\documentclass{osa-article}

\journal{oe}
\articletype{Research Article}
\usepackage{cancel}
\definecolor{Ogg}{rgb}{0.0, 0.5, 0.0}
\definecolor{Brg}{rgb}{0.0, 0.26, 0.15}

\definecolor{DrkGrn}{RGB}{0,0.667,0}
\usepackage{subcaption}
\usepackage{tikz}

\usetikzlibrary{decorations.markings}
\tikzset{->-/.style={decoration={
  markings,
  mark=at position #1 with {\arrow{>}}},postaction={decorate}}}
\tikzset{-<-/.style={decoration={
  markings,
  mark=at position #1 with {\arrow{<}}},postaction={decorate}}}
\usetikzlibrary{external}
\def\Re{{\textup{Re}}}

\def\innerprod(#1,#2){{\left<#1\,,\,#2\right>}}
\def\Set#1{{\left\{#1\right\}}}
\def\qquadtext#1{\qquad\textup{#1}\qquad}
\def\qquadand{\qquadtext{and}}

\def\dfrac#1#2{\frac{d #1}{d #2}}

\usepackage{fancyhdr}
\fancypagestyle{firststyle}
{
   \fancyhf{}
   
\fancyfoot[C]{1}
\fancyfoot[R]{\footnotesize\bf File: \jobname.tex}
}
\def\VEft{{\tilde{\boldsymbol E}}}
\def\VBft{{\tilde{\boldsymbol B}}}

\def\Vzero{{\boldsymbol 0}}
\def\Vi{{\boldsymbol e}_x}
\def\Vj{{\boldsymbol e}_y}
\def\Vk{{\boldsymbol e}_z}

\def\fsc{{\hat\omega}}

\def\fscC{\fsc_{{\textup{c}}}}

\def\ErrMax{\boldsymbol{\cal E}_{\textup{Max}}}
\def\ErrBdd{\boldsymbol{\cal E}_{\textup{bdd}}}

\def\lamy{\kappa_y}
\def\lamz{\eta}
\def\MatPF{{\Psi}}
\def\Lz{L_z}
\def\Lx{{\cal L}_x}
\def\Ly{L_y}
\def\Lnul{L_0}

\def\kscale{{\hat k}}

\def\forcePeriod{{f_{\textup{P}}}}

\def\px{p_x}
\def\py{p_y}
\def\TM{{\textup{TM}}}

\def\mmeters{{\textup{mm}}}

\def\MatPF{{\Psi}}
\def\Lz{L_z}
\def\Lx{{\cal L}_x}
\def\Ly{L_y}
\def\Lnul{L_0}
\def\kapx{\kappa_x}

\definecolor{blueColPal}{rgb}{0.392157, 0.560784, 1.}
\definecolor{orangeColPal}{rgb}{0.996078, 0.380392, 0.}
\definecolor{magentaColPal}{rgb}{0.862745, 0.14902, 0.498039}

\begin{document}

\title{Broadband Terahertz Generation in a Corrugated Waveguide with matched Phase and Group Velocities}

\author{Sergey S. Siaber\authormark{1,2}, Jonathan Gratus\authormark{1,2,*},  Rebecca Seviour\authormark{3},  Steven P. Jamison\authormark{1,2} and Taylor Boyd\authormark{1,2}}

\address{
\authormark{1}Dept Physics, Lancaster University, Lancaster, UK\\
\authormark{2}Cockcroft Institute of accelerator science,  Daresbury, Warrington,
WA4 4AD, UK\\
\authormark{3}Ion Beam Centre, University of Huddersfield, Huddersfield, UK}

\email{\authormark{*}j.gratus@lancaster.ac.uk} %% email address is required

% \tableofcontents



\begin{abstract}
We show that it is possible to design corrugated waveguides where phase and group velocities coincide at an inflection point of the dispersion relation, allowing an extended regime of interaction with a charge particle beam. This provides a basis for designing travelling slow-wave structures with a broadband interaction between  relativistic charged particle beams and propagating terahertz waves allowing an energy exchange between beam and wave, amplifying terahertz radiation.

We employ Fourier-Mathieu expansion, which gives approximate analytic solutions to Maxwell equations in a corrugated waveguide with periodically undulating cross-section. Being analytic, this enables quick design of corrugated waveguides, determined from desirable dispersion relations.

 We design a three dimensional waveguide with the desired dispersion and confirm the analytical predictions of the wave profile, using numerical simulations. 
Madey's theorem is used to analyse the strength of the wave-beam interaction, showing that there is a broad frequency interaction region.
\end{abstract}

%%%%%%%%%%%%%%%%%%%%%%%%%%  body  %%%%%%%%%%%%%%%%%%%%%%%%%%
\section{Introduction}
%%%%%%%%%%%%%%%%%%%%%%%%%%%%%%Sub out RF
\label{ch_Intro}

Driven by the growning demand of applications, from material science to telecommunications, from biology to biomedicine, recent years have seen a rapid rise in the development of coherent terahertz (THz) sources.
Technologies used to generate THz radiation include laser-driven emitters, solid state oscillators, gas and quantum cascade lasers.  Laser driven emitters, the most widely used sources of pulsed THz radiation, are based on frequency down-conversion from the optical region. THz pulses energies exceeding 1\,mJ and 10's MW peak power have been demonstrated in period poled nonlinear sources driven by high energy near-IR ultrafast lasers\cite{Lemery2020}, while more main stream laser systems are capable of 10's of $\mu$J energies and efficiencies in region of 0.1\% \cite{Mosley2023}. In solid state oscillators the transit time of carriers through semiconductor junctions limits the frequency and power that can be generated, generating around 100 mW at 100 GHz, but the power falls off as $f^{-2}$\cite{Eis}. Optically pumped gas lasers are the oldest technologies for THz generation, generating between 0.3 to 5 THz at around 100mW \cite{mat}.  Quantum cascade lasers are a relatively new approach to generating THz radiation generating between 1 to 4 THz at mW power levels \cite{fatt}.

We consider an electron beam driven approach to THz generation. Electron beam approaches are either non-relativistic/moderately relativistic vacuum electronics devices (VED) or ultra-relativistic accelerator based radiators, where a modulated electron beam generates THz via transition, Cherenkov, Smith-Purcell, or undulator radiation. We present a moderately relativistic VED approach that uses a novel corrugated metallic waveguide as a slow-wave structure to generate THz radiation via coherent spontaneous emission. More broadly metallic waveguides with corrugations or modulations on the metallic boundary are of wide interest for their ability to tune the EM propagation characteristics through the corrugation structure. Corrugations in the form of rectangular groves orthogonal to the propagation axis and sub-wavelength separation form slow-wave structures that give rise to dispersion relations similar to that of a dielectric lined waveguides with the groove geometry and spacing determine the effective dielectric properties. In such an arrangement the dispersion relation can be obtained through a continuing sequence of mode-matching along the waveguide with step-wise changes in cross-section. The mode-matching over the repeating structure leads to a global eigenvalue problem from which the dispersion relation is obtained.

%%%%%%%%%%%%%%%%%%%%%%%%%%%%%%%%%%%%%%%%%%%%%%%%%%%%%%%%%%%%%%%%%%%%%%
%               FIGURE 1
%%%%%%%%%%%%%%%%%%%%%%%%%%%%%%%%%%%%%%%%%%%%%%%%%%%%%%%%%%%%%%%%%%%%%%

\begin{figure}[htbp]
\centering\includegraphics[width=9cm]{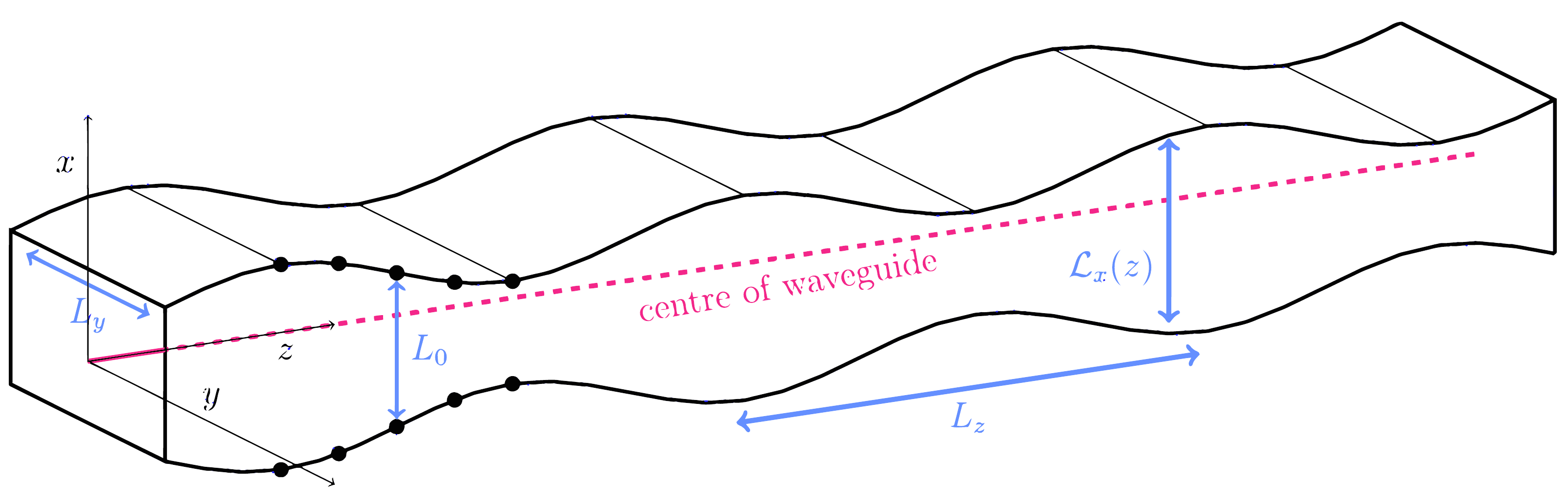}
\caption{Rectangular Corrugated Waveguide, with periodically undulating cross-section. It is designed to vary in such a way that phase and group velocities are very close in an extended frequency range.}
\label{fig_corr_waveguide}
\end{figure}

In this paper, we use a waveguide which has smooth sinusoidal undulations in the waveguide cross-section, its form is shown in figure \ref{fig_corr_waveguide}. This waveguide, has symmetric  undulations in cross-section height, given by the function $\Lx(z)$, unlike sine-waveguide, considered in \cite{zhang,xiaogen}, where cross-section shape and dimensions are constant, but its centre is oscillating. For this structure we can find explicit approximate solutions for the EM fields. The geometry of the waveguide structure is parametrised by four parameters. Two of these are traverse dimensions, one, denoted as $\Lx(z)$, is varied along the waveguide and oscillates around the average value, marked as $L_0$, and the other is fixed (denoted as $\Ly$). The other two are the spatial period of corrugation, $\Lz$,   and depth of the corrugations $q$. Theoretical technique that predicts EM propagating mode in an analytic form enables us to quickly scan this parameter space for desired dispersion and EM field properties.  We applied this approach to  find the parameters where group and phase velocities coincide,  and do so at a point of inflection of the dispersion relation. Such a point corresponds to  particle-wave phase matching at a broad range of frequencies. We call this point the {\em coincident inflection point} (CIP). Research presented in this manuscript originated from the ideas of EM waves propagation in corrugated wire media\cite{gratus2015,boyd2018Cust,boyd2018Mode}.

The concept of engineering the dispersion of VEDs using a corrugated waveguide is not new. 
For example, some gyrotron amplifier designs utilise corrugated waveguides\cite{Bratman2000} to engineer dispersion, 
matching phase velocity with that of an electron beam, 
using a longitudinal sinusoidal modulation in a cylindrical waveguide wall together with a 3-fold helical twist in the modulation. This dispersion relation can be obtained from a perturbative coupling between uniform cross-section waveguide guide modes, with an anti-crossing between mode dispersion arising from the coupling~\cite{Cook1998, Denisov2000}. While gyrotron structures achieve the desired dispersion, matching  wave-particle velocity's, they also require a complex electron beam structure and helical propagation in high magnetic fields; the electron beam must have annular and helical propagation matching to couple to the EM mode ~\cite{nusinovich2004introduction}.

 More recently researchers have considered uniform rectangular cross-section waveguides to engineer the dispersion in VEDs, 
such as the piecewise sine waveguide \cite{zhang} for high-power terahertz (THz) travelling wave tubes (TWTs). This topology results in multiple modes existing simultaneously point in the dispersion curve (see figure 4 in Zhang \cite{zhang}), which can result in mode competition and noise on the generated RF.
 Other researchers have considered a uniform cross-section dielectric lined waveguides have found application in supporting electron acceleration driven by terahertz-frequency laser pulses~\cite{Nanni2015, Zhao2019, Wong2013, Hibberd2020}. The dielectric layers allow for phase-matching of the drive laser (at THz frequencies) $LSM_{10}$ modes with sub-relativistic electron beams. However, unlike in the helical gyrotron structures, the phase-velocity matching intrinsically comes with a group-velocity mismatch, and the temporal walk-off between the EM pulse and electron beam is a limiting factor in application of these dielectric lined waveguides to particle acceleration.

In this article,  we present the design of a rectangular corrugated, with periodically undulating cross-section, waveguide that enables efficient THz generation through a number of dispersion relation characteristics. Achieving these characteristics in this waveguide is enabled by our analytical technique. We seek and find, a structure that has a dispersion that provides both group and phase-velocity matching like a gyrotron, yet has the potential to support axial beams like uniform cross section waveguides, without mode competition.
Our approach provides a perturbative solution to the waveguide modes, from which we show that the dispersion relation is a solution to Mathieu's Floquet equation with two free parameters: a cut-off frequency $\fscC$, determined by the relative dimensions $\Lnul/\Lz$ and $\Ly/\Lz$ of the waveguide, and $q$ which determines the relative height of the undulations. Solution of the associated eigenvalue problem provides the dispersion relation, and the Floquet coefficients describing the longitudinal spatial-harmonics of the propagating mode. Solutions with group and phase velocity simultaneously being equal, and subluminal, are shown to exist.
This solution can be used for the coupling system as well as the beam/wave waveguide interaction region.

Unlike previous approaches, in this paper we determine the waveguide profile from the desired dispersion relation and not set a priori, 
i.e. we determine a family of waveguide geometries that satisfy the  dispersion relation. Solutions are of the form of a slowly varying sinusoidal-like longitudinal modulation in the waveguide cross-section, and an EM mode with a longitudinal electric field  on axis. 
In addition to matching the phase velocity and electron beam velocity,  to maximise energy transfer between the electron beam and EM wave we require the spatial components of the electric field of the EM wave to be parallel with the  beam propagation.

The paper is structured as follows. In section \ref{ch_Sol} we give the explicit
approximate EM modes for the corrugated
waveguide, based on Mathieu's functions. We then demonstrate, in section \ref{ch_CIP}, a method of finding the appropriate parameters $\fscC$ and $q$, which we apply this to our goal of finding the CIP. In section \ref{ch_Siml}, we use numerical CST Microwave studio simulations confirm the accuracy of the analytic approximation and to find how the CIP moves slightly when going from this approximation to numerical simulations. In section \ref{ch_Force} we show the force that a charged particle would experience, and apply Madey's theorem \cite{LuchiniMotzFEL,MadeyOrig,yan,sev1} to estimate THz energy being emitted as a result of particle--wave interaction in the waveguide.  We conclude by discussing future directions. In addition, in the Supplemental Document we show how to construct the other EM modes, analyse further Mathieu's equation and how it leads to interaction zones, perform error analysis on the approximate solutions, show that there are no subluminal modes in the first interaction zone, and expand on the subject of simulating numerical field patterns and comparing it to analytical field patterns.

%%%%%%%%%%%%%%%%%%%%%%%%%%%%%%%%%%%%%%%%%%%%%%%%%%%%%%%%%%%%%%%%%%%%%%

\section{Explicit Solution for a corrugated waveguide profile}
\label{ch_Sol}

We consider a rectangular corrugated waveguide shown in figure
\ref{fig_corr_waveguide}.  We assume it has two flat walls set a $\Ly$
distance  apart. The other two walls have a symmetric undulating profile
 a distance $\Lx(z)/2$ from the centre, i.e. the two walls are a distance $\Lx(z)$ apart. The cross section
dimensions are given by the constant width $\Ly$ and the variable
height $\Lx(z)$ which depends on the positions $z$ along the
waveguide. We choose the $(x,y)$
coordinates to be $(0,0)$ in the centre of the waveguide so that the
faces of the waveguides are at $y=\pm \Ly/2$ and $x=\pm \Lx(z)/2$.
The undulations in $\Lx(z)$ are periodic, with a period $\Lz$, oscillating around $\Lnul$ value.
The exact form of undulating height $\Lx(z)$ of corrugations will be given below.

Solving Maxwell's vacuum equations in the  frequency domain (with factor $e^{-i\omega t}$)  
%Our solution is an approximation so it does not solve Maxwell's equations exactly, instead we solve the following
%[
\begin{align}
&\nabla\times\VBft + i\omega\,c^{-2}{\VEft} = \Vzero,%\nabla\cdot\VEft=0,
%\label{Sol_Max_Apprx_Gauss},
%\\
&&\nabla\cdot\VBft=0,%1\nabla\times\VBft + i\omega\,c^{-2}{\VEft} = 0,%\nabla\cdot\VBft=0,
\label{Sol_Max_Apprx_NoMon}
\\
&\nabla\times\VEft - i\omega{\VBft} = \ErrMax
,
%\label{Sol_Max_Apprx_Fara}
%\\
&&\nabla\cdot\VEft=0.%1\nabla\times\VBft + i\omega\,c^{-2}{\VEft} = 0,
\label{Sol_Max_Apprx_Amp}
\end{align}
%]
together with the boundary conditions
%[
\begin{align}
&\VEft_\parallel |_{\textup{Bdd}} = \ErrBdd,
%\label{Sol_Max_Bdd_EPll}
%%\\
&\VBft_\perp |_{\textup{Bdd}} = \Vzero,
\label{Sol_Max_Bdd_BPerp}
\end{align}
%]
where the errors $\ErrMax$ and $\ErrBdd$ for our waveguide, are small.
Our solution is thus an approximation so it does not solve Maxwell's equations exactly. 
%Our solution is an approximation so it does not solve Maxwell's equations exactly.

We consider $\TM_{\px\py}$ modes, with positive odd integers $\px$ and $\py$, as these are the modes with a longitudinal component of $\VEft$ in the centre of the waveguide that we require. 
%Guided by the expectation of a $\TM_{\px\py}$ like mode, where %$\px$
%and $\py$ are odd integers,
We start by specifying the ansatz for the
magnetic field $\VBft$,
%[
\begin{align}
\VBft &= B_0\, c^{-2}\,(-i \omega)\,\phi(\lamz\,z)
\big(
\lamy\cos(\kapx\,x)\,\sin(\lamy\,y)\,\Vi
-
\kapx\,\sin(\kapx\,x)\,\cos(\lamy\,y)\,\Vj
\big)
\label{Sol_B_flat}
\end{align}
%]
where $\Set{\Vi,\Vj,\Vk}$ are the unit vectors along the  axes,
%[
\begin{align}
\kapx(z)=
\frac{\pi\,\px}{\Lx(z)}
,\quad %\label{Sol_alpha}
\lamy 
&= 
\frac{\pi\,\py}{\Ly}
%\label{Sol_lambda_y}
%\qquad
, \,\,\textrm{and} \quad
\lamz =\frac{\pi}{\Lz}.
\label{Sol_lambda_z}
\end{align}
%]
Trivially, $\VBft$ satisfies divergence equation in (\ref{Sol_Max_Apprx_NoMon}). Substituting  $\VBft$ into curl equations and combining these into one equation yields the 2nd order ODE for $\phi(\lamz\,z)$
%[
\begin{align}
\phi''(\lamz\,z) + 
\lamz^{-2} \,\big(c^{-2}\omega^2-\kapx^2-\lamy^2\big)\,\phi(\lamz\,z)
&=
0.
\label{Sol_phi_eqn_alpha}
\end{align}
%]
From Gauss's equation (\ref{Sol_Max_Apprx_NoMon}) the electric field $\VEft$ is
%[
\begin{equation}
\begin{aligned}
\VEft
&=
 B_0\,
\Big(\big(\kapx'\,\sin(\kapx\,x)+\kapx'\kapx\,x\,\cos(\kapx\,x)\big)
\phi(\lamz\,z)
+
\lamz\kapx\,\sin(\kapx\,x)\,\phi'(\lamz\,z)
\Big)
\cos(\lamy\,y)\,\Vi
\\&\qquad\qquad\qquad
+ 
B_0\,\lamy\Big(
-
\kapx'\,x\,\sin(\kapx\,x)\,\phi(\lamz\,z)
+
\lamz \cos(\kapx\,x)\,\phi'(\lamz\,z)
\Big)
\sin(\lamy\,y)\,\Vj
\\&\qquad\qquad\qquad
- 
B_0\,
(\kapx^2+\lamy^2)
\,\cos(\kapx\,x)\,\cos(\lamy\,y)\phi(\lamz\,z) 
\Vk
\end{aligned}
\label{Sol_E}
\end{equation}
%]
We see  $\VBft$, $\VEft$, given by eqs. (\ref{Sol_B_flat}) and (\ref{Sol_E}), automatically satisfy divergence equation in 
(\ref{Sol_Max_Apprx_Amp}) since
$\nabla\cdot\VEft=\nabla\cdot\big(c^2(i\omega^{-1})\nabla\times\VBft\big)=0$.
In Supplemental Document, in the section SD2, we give the formula for the error
$\ErrMax$. We also calculate the error
$\ErrBdd$ in the boundary, and we show that these errors are small if waveguide profile have a shallow gradient.  We note that in the centre of the waveguide, where
$x=0$, then $\ErrMax|_{x=0}=0$.  The general solution to Maxwell's equations for our waveguide is given by the sum of TM and TE modes. These are briefly discussed in the supplementary document, section SD4.
We presume that most of the electrons are travelling along the centre of the waveguide, where solution is more accurate.

Equation (\ref{Sol_phi_eqn_alpha}) can be rewritten in the form of Mathieu equation (here we defined scaled variable $\zeta := \pi z/\Lz  = \lamz\,z$ )
%[
\begin{align}
\phi''(\zeta) + 
\big(a - 2 q\cos(2\zeta)\big) \phi(\zeta)
=0.
\label{Mathieu_Eq}
\end{align}
%]
Indeed, by stipulating an identity between corresponding terms in (\ref{Mathieu_Eq}) and (\ref{Sol_phi_eqn_alpha}), and substituting explicit form of $\lamz$, $\lamy$ and $\kapx$, we get
%[
\begin{equation}
\begin{aligned}
a - 2 q\!\cdot\!\cos(2\zeta)
&\equiv
\lamz^{-2} \,\big(c^{-2}\omega^2-\kapx^2-\lamy^2\big)
\\&=
\Lz^{2} \,\big(c^{-2}\pi^{-2}\omega^2-{\px^2}\Lx(z)^{-2}-\py^2\Ly^{-2}\big)
\\&=
\Lz^{2}
\,\big(c^{-2}\pi^{-2}\omega^2-{\px^2}\Lnul^{-2}-\py^2\Ly^{-2}\big)
-
\Lz^{2} \,\big({\px^2}\Lx(z)^{-2}-{\px^2}\Lnul^{-2}\big),
\label{AnadQEquiv}
\end{aligned}
\end{equation}
%]

\noindent where we choose the separation into the  dimensionless Mathieu parameters $a$ and $q$ as

%[
\begin{align}
a = \Lz^{2}
\,\big(c^{-2}\pi^{-2}\omega^2-{\px^2}\Lnul^{-2}-\py^2\Ly^{-2}\big),
\label{Match_a_res}
\end{align}
\begin{align}
2q\cos(2\zeta) = \Lz^{2} \,\big({\px^2}\Lx(z)^{-2}-{\px^2}\Lnul^{-2}\big),
\end{align}

%]
\noindent In the following we choose a constant $q$, which  defines the wave-guide profile 
\color{black}

%[
\begin{align}
\Lx(\zeta) = 
\Big(\Lnul^{-2} + 
2\Lz^{-2}\,\px^{-2}\, q\,\cos(2\zeta) \Big)^{-1/2}
\label{Match_Lx_L0}
\end{align}
%]
Under the assumption that $q\,\Lnul^2\,\Lz^{-2}$ is small,
waveguide profile $\Lx$  takes simply a form of  sinusoidally undulating function
%[
\begin{align}
\Lx(\zeta)\approx \Lnul
- 
\Lnul^3 \, \Lz^{-2}\, q\,\cos(2\zeta)
\label{L_Approx_Form}
\end{align}
%]
It this form, it is evident that $\Lnul$ is the average height of
the waveguide and parameter $q$ determines the height of the corrugations. 
For the example waveguides to be discussed here we will choose $q=0.1$ and $\Lnul/\Lz\approx 1$.

Solutions of  the equation (\ref{Mathieu_Eq}) are the Mathieu special
functions. These solutions, as stated in Floquet's Theorem \cite{Eastham1973TheST}, can be presented in the form
%[
\begin{align}
\phi_{a,q}(\zeta) = e^{i\kscale \zeta}\,\MatPF_{a,q,\kscale}(\zeta),
\label{Mat_Floque}
\end{align}
%]
where $\MatPF_{a,q,\kscale}(\zeta)$ is periodic,
%[
\begin{align}
\MatPF_{a,q,\kscale}(\zeta+\pi) = \MatPF_{a,q,\kscale}(\zeta),
\label{Mat_P_Periodic}
\end{align}
%]
where we have introduced the subscripts on $\phi_{a,q}(\zeta)$ to make explicit the dependencies.  
$\kscale$ is the Mathieu exponent, with a value that is determined by $a$ and $q$. In this form we see that $\kscale$ may be considered as a dimensionless wavevector describing the wave propagation, related to the wavevector by
$k = \lamz\kscale=\pi\kscale/L_z $.   
The relationship between $k$ and $\omega$, or equivalently between $k$ and $a$, therefore defines the dispersion relationship for the traveling wave solutions to equations~(\ref{Sol_phi_eqn_alpha}) and (\ref{Mathieu_Eq}), while $\MatPF_{a,q,\kscale}(\zeta)$ provides a structure function for the field variation within a single period or cell of the undulating waveguide.

%Given the quantities $a$ and $q$ there is a formula for $\kscale$. Since $a$ depends on $\omega$, then for a fixed $q$ we can find the dispersion relationship between $k$ and $\omega$. For example, see figure \ref{fig_Mat-Force_Disp}.  
%Observe that for a particular $q$, if $(\omega,\kscale)$ is on the dispersion relation so are $(\omega,\kscale+2n)$ and $(\omega,2n-\kscale)$ for all integers $n$. The details are given in Supplemental Document, in the section SD2. This gives rise to the separate zones which is important for considering the wave-particle interaction, as discussed in section \ref{ch_Force}.
\color{black}

%%%%%%%%%%%%%%%%%%%%%%%%%%%%%%%%%%%%%%%%%%%%%%%%%%%%%%%%%%%%%%%%%%%%%%
%               FIGURE 2 a
%%%%%%%%%%%%%%%%%%%%%%%%%%%%%%%%%%%%%%%%%%%%%%%%%%%%%%%%%%%%%%%%%%%%%%

\begin{figure}
\centering
\begin{tikzpicture}[scale=1.6]
\draw(0,0) node[above right=-8pt] {\includegraphics[width=0.99\textwidth]{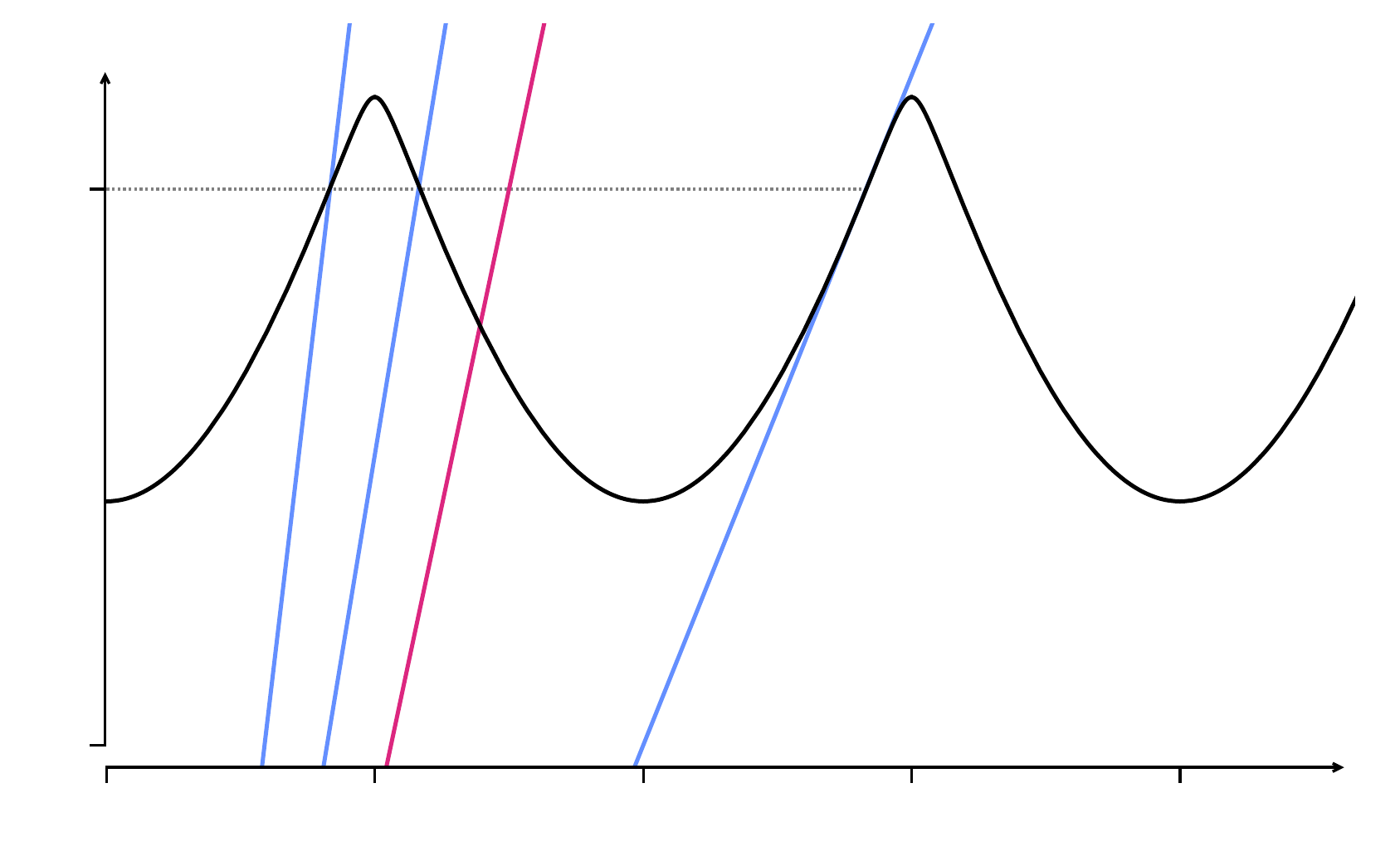}} ;
%Y-AXeS LABELS/TICKS
\draw(0.15,4.4) node{\Large$\fsc$} ;
\draw(0.275,3.85) node{\normalsize 1.5} ;
\draw(0.22,1.99) node{\normalsize 1.255} ;
\draw(0.2,0.58) node{\normalsize 1.06} ;
%X-AXeS LABELS/TICKS
\draw(7.6,0.65) node{\Large$\kscale$} ;
\draw(0.54,0.18) node{\normalsize 0} ;
\draw(2.12,0.18) node{\normalsize 1} ;
\draw(3.7,0.18) node{\normalsize 2} ;
\draw(5.28,0.18) node{\normalsize 3} ;
\draw(6.86,0.18) node{\normalsize 4} ;
%LINE LABELS
\draw (1.75,1.9) node[rotate=82.5] 
  {\small superluminal particle, $1^{\text{st}}$ zone} ;
\draw (2.19,1.9) node[rotate=79] %\draw (1.9,1.9) node[rotate=80] 
  {\small superluminal particle, $2^{\text{nd}}$ zone} ;
\draw (2.6,1.69) node[rotate=78]%,green!60!black] 
  {\small \textcolor{magentaColPal}{lightline}} ;
\draw (4.305,1.69) node[rotate=68] 
  {\small subluminal particle, $3^{\text{rd}}$ zone} ;
\end{tikzpicture}
\caption{Corrugated waveguide dispersion relation (blue line). The light
  line is shown in pink and the three black lines represent three
  particle beam velocities that would interact with monochromatic EM wave in: the first zone (corresponding hypothetical particle beam velocity would have been $1.8c$), second zone (similarly, hypothetical $\beta_0$ would have been $1.29$), and third zone where matching velocity is subluminal (particle beam $\beta_0=0.53$).}
\label{fig_Mat-Force_Disp} 
\end{figure}
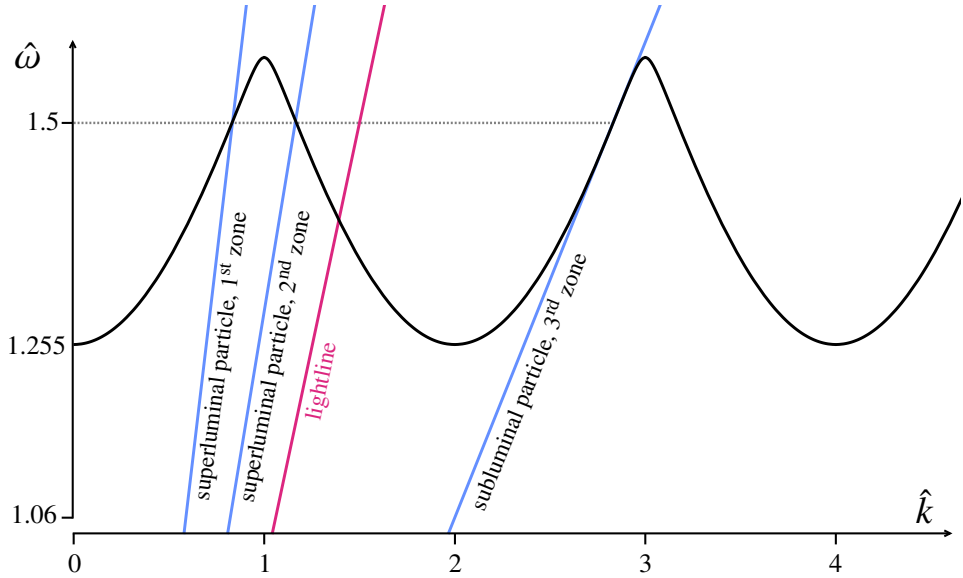
%%%%%%%%%%%%%%%%%%%%%%%%%%%%%%%%%%%%%%%%%%%%%%%%%%%%%%%%%%%%%%%%%%%%%%
%               
%%%%%%%%%%%%%%%%%%%%%%%%%%%%%%%%%%%%%%%%%%%%%%%%%%%%%%%%%%%%%%%%%%%%%%

%%%%%%%%%%%%%%%%%%%%%%%%%%%%%%%%%%%%%%%%%%%%%%%%%%%%%%%%%%%%%%%%%%%%%%

%\color{blue}
% \subsection{Dispersion relation}
% \label{ch_Scal}

The shape of the corrugated waveguide is determined by four parameters: the length of one period of the corrugation $\Lz$, the width of the waveguide $\Ly$, the separation of the corrugated surfaces at the mid point of the corrugation $\Lnul$ and $q$ a parameter related to the depth of the corrugation. 
Once these four parameters are chosen this defines a dispersion relation between the angular frequency $\omega$ and the wavevector $k$ for the corresponding EM wave.  In practice, rather than specifying a frequency $\omega$, or  equivalently $a$, and then evaluating the corresponding $\kscale$, we perform the calculation in reverse.
Specifying $\kscale$ and the structure geometry the corresponding value of $a$ can be determined from an eigenvalue evaluation \cite{ArscottMathiueEq}.

An example of the dispersion relation obtained for $q=0.1$ is shown in Figure~\ref{fig_Mat-Force_Disp}. To highlight the generality we have displayed dispersion relation with the following dimensionless parameters,
%[

\begin{align}
\fsc &= \frac{\Lz}{\pi\,c}\,\omega
 \qquadand
\fscC = \bigg(\frac{\Lz^2}{\Lnul^2} + \frac{\Lz^2}{\Ly^2}\bigg)^{1/2}.
\label{Scal_Scaling}
\end{align}
so that 
equation (\ref{Match_a_res}) the Mathieu parameter $a$ becomes
%[
\begin{align}
a_{\kscale,q} = \fsc^2 - \fscC^2
\label{Scal_a}
\end{align}
%]
%]
Thus,  the number of parameters for the dispersion relation in the waveguide reduces to just two, 
$\fscC$ and $q$. For small values of $q$, such as we are considering here, $\fscC$ may be regarded as the normalised cut-off frequency.
Given these two parameters we can quickly generate the scaled dispersion relation for $\fsc$ and $\kscale$.

One of the main advantages of this approach is that we can quickly scan the parameter space in order to find a desirable dispersion  relationship. A wide range of behavior can be found through the parameterisation of equations~(\ref{Scal_Scaling}-\ref{Scal_a}) and a numerically straightforward eigenvalue evaluation of $a_{\kscale,q}$.

\section{Particle-wave synchronism.}
\label{ch_CIP}

\color{black}

%$c\mkern-5mu\cdot\mkern-4mu\beta_e$

For extended wave-particle interaction it is necessary to find propagating EM mode in which charged particle and electromagetic phase wave remain in phase, with particle ($c\beta_e$) and phase-velocity ($c{\kscale}/{\fsc}$) being equal. We aim to find more stringent solutions where, for ceratin $\kscale, \fsc(\kscale), \fscC, q$ values, the the phase-velocity  and group velocity both match the particle velocity
%
% \begin{align}
% \frac{\fsc\left(\kscale,\fscC,q\right)}{\kscale}=\frac{\partial\fsc\left(\kscale,\fscC,q\right)}{\partial\kscale}=\beta_e
% \label{PhaseGroupMatch_explicitly}
% \end{align}
%
\begin{align}
\frac{\fsc}{\kscale}=\dfrac{\fsc}{\kscale}=\beta_e
\label{PhaseGroupMatch_explicitly_alt}
\end{align}
where $\fsc(\kscale)$ depends on the values of the parameters $\fscC$ and $q$. In order for the interaction to occur in a range of frequencies (as opposed to a single frequency), we seek to simultaneously satisfy a third constraint: zero group velocity dispersion (GVD), formulated as having an inflection point on the dispersion when phase and group velocities are matched,  i.e. 
%[]
\begin{align}
\frac{\fsc}{\kscale}=\dfrac{\fsc}{\kscale} \quad  \textrm{and}  \quad \frac{d^2 \fsc}{d \kscale^2}=0
\label{CIP_explicitly}
\end{align}
%]
Such EM mode maintains in-phase interaction, is free of EM pulse walk-off, and minimizes the pulse dispersion within the waveguide.
We refer to frequencies and wavenumbers that satisfy these conditions as the coincident inflection point (CIP). We regard, the CIP as an important design and operating point that enables enhanced broadband interaction with an electron beam, and provided initial synchronicity, continuous energy transfer from beam to wave or from wave to beam.  

In the dispersion diagram of figure \ref{fig_Mat-Force_Disp} we show the $\fsc$-$\kscale$ relationship for particles that would be phase-synchronised at normalized frequency of $\fsc = 1.5$ (this corresponds to $f \simeq 474\,$GHz in the example structure described in detail later), and for normalised cut-off  frequency  $\fscC=1.255$ and corrugation parameter $q=0.1$.

The first zone corresponds to a particle traversing a single cell in one oscillation period; the 2nd zone to two oscillations per cell traversal, and the 3rd zone to three oscillations per traversal.
Synchronism in the first zone is generally as it requires a superluminal particle, as is also the case for the 2nd zone in this example.
For the 3rd zone however, synchronism can be obtained for particle velocity of $\beta_e = 0.53$ (corresponding approximately 92\,keV electrons).
The intercept of the particle and wave dispersion lines (matched phase-velocity), with parallel tangents (matched group velocity), occurs at the point of inflection in the wave dispersion (zero GVD). It is therefore represents a CIP for $\beta_e = 0.53$, and $\fsc = 1.5, \fscC=1.255, q=0.1$.

% \textcolor{red}{?}In figure~\ref{fig_Inf_group_phase_vel} we show the group and phase velocities in the region of the CIP of figure \ref{fig_Mat-Force_Disp}, highlighting the coincidence in group and phase velocities and the velocity inflection point.
% For comparison, the behavior of a uniformly rectangular wave-guide is also shown.\textcolor{red}{?}

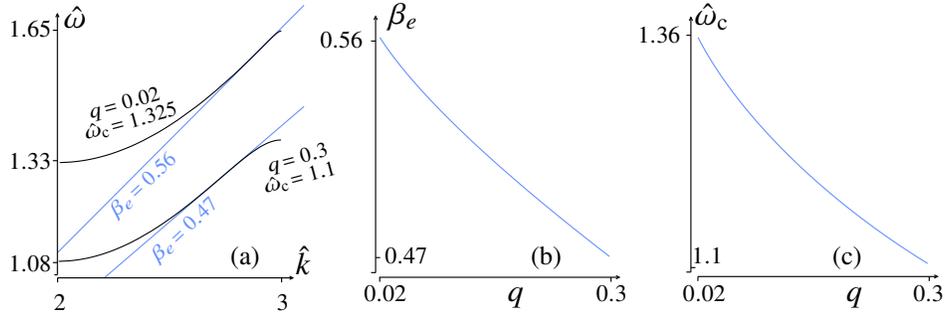
\begin{figure}[htbp]
    \centering
\begin{tikzpicture}[scale=1]
\draw (0,0) node[above right=-.0005cm]{
    \includegraphics[width=0.95\textwidth]{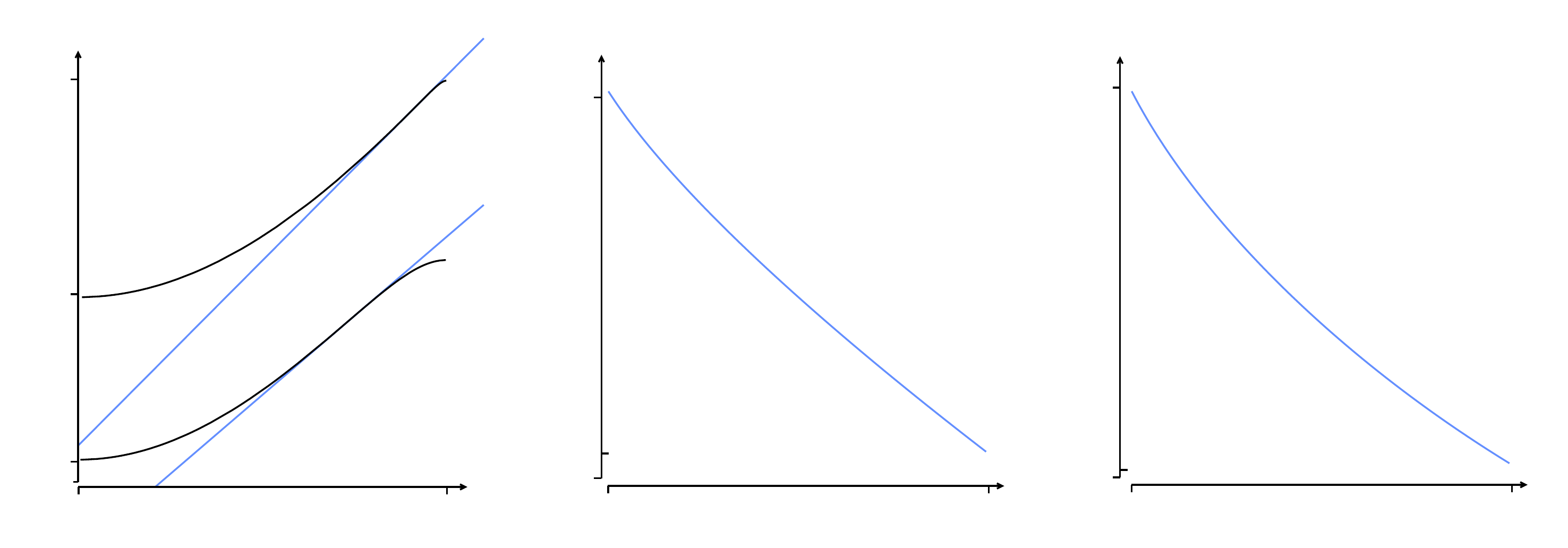}
};
%1ST SUBfigure
\draw(0.99,4.09) node{\large$\fsc$} ;
\draw(4.025,0.85) node{\large$\kscale$} ;
\draw (0.39,3.93) node {\small $1.65$} ;
\draw (0.39,2.17) node {\small $1.33$} ;
\draw (0.39,0.81) node {\small $1.08$} ;
\draw (0.79,0.3) node {\small $2$} ;
\draw (3.75,0.3) node {\small $3$} ;
%LINE LABELS
\draw (1.65,2.93) node[rotate=15] 
  {\footnotesize  $q\!=0.02$} ;
  \draw (1.72,2.69) node[rotate=15] 
  {\footnotesize $\fscC\!=1.325$} ;
  \draw (3.95,2.2) node[rotate=15] 
  {\footnotesize  $q\!=0.3$} ;
  \draw (3.97,1.94) node[rotate=15] 
  {\footnotesize $\fscC\!=1.1$} ;
  \draw (1.9,1.85) node[rotate=43] 
  {\footnotesize \textcolor{blueColPal}{$\beta_e\!=0.56$}} ;
\draw (2.45,1.25) node[rotate=43] 
  {\footnotesize \textcolor{blueColPal}{$\beta_e\!=0.47$}} ;
%2ND SUBfigure
\draw(5.35,4.09) node{\large$\beta_e$} ;
\draw(6.85,0.35) node{\large$q$} ;
\draw (4.53,3.78) node {\small $0.56$} ;
\draw (5.39,0.91) node {\small $0.47$} ;
\draw (5.11,0.41) node {\small $0.02$} ;
\draw (8.12,0.41) node {\small $0.3$} ;
%3RD SUBfigure
%\draw(9.45,4.09) node{\large$\beta_e$} ;
\draw(9.45,4.09) node{\large$\fscC$} ;
%\draw(11.35,0.85) node{\large$\fscC$} ;
\draw(11.35,0.35) node{\large$q$} ;
%\draw (8.75,3.78) node {\small $0.56$} ;
\draw (8.75,3.85) node {\small $1.36$} ;
%\draw (8.82,0.91) node {\small $0.47$} ;
\draw (9.39,0.91) node {\small $1.1$} ;
%\draw (9.25,0.41) node {\small $1.1$} ;
\draw (9.35,0.41) node {\small $0.02$} ;
%\draw (12.1,0.41) node {\small $1.36$} ;
\draw (12.33,0.41) node {\small $0.3$} ;
%SUBFIGURE LETTERS
\draw(3.25,0.88) node{\normalsize (a)} ;
\draw(7.25,0.88) node{\normalsize (b)} ;
\draw(11.25,0.88) node{\normalsize (c)} ;
\end{tikzpicture}
    \caption{End of range dispersion relations (black lines) and CIP particle lines (blue lines) are shown on subfigure (a). CIP velocity dependence on $q$ values is shown on subfigure (b). Relation between CIP parameters $q$ and $\fscC$ is given on subfigure (c).}
    \label{fig_Inf_v_q}
\end{figure}

More generally, for a given waveguide undulation scale parameter $q$, we find, one unique CIP. That is, for a given $q$, we can find the corresponding unique values of $\fsc$, $\kscale$ and $\fscC$ for the CIP. For a $q$ range from 0.0 to 0.3, corresponding CIP synchronous particle velocity extends from 0.56$c$ to 0.47$c$. Third zone of the dispersion relation for the ends of this range are shown on Fig.\ref{fig_Inf_v_q} (a) toghether with synchronous particle lines. Correspondence between particle velocities and parameters $q$ and $\fscC$ for which CIP occurs is shown on Fig.\ref{fig_Inf_v_q} (b) and (c) accordingly. We also observe that for $q\to 0$, there is an absolute maximum velocity for the CIP, $v=0.5754c$.

%%%%%%%%%%%%%%%%%%%%%%%%%%%%%%%%%%%%%%%%%%%%%%%%%%%%%%%%%%%%%%%%%%%%%%
%         
%%%%%%%%%%%%%%%%%%%%%%%%%%%%%%%%%%%%%%%%%%%%%%%%%%%%%%%%%%%%%%%%%%%%%%

%%%%%%%%%%%%%%%%%%%%%%%%%%%%%%%%%%%%%%%%%%%%%%%%%%%%%%%%%%%%%%%%%%%%%%
\section{Numerical simulation}
\label{ch_Siml}
%%%%%%%%%%%%%%%%%%%%%%%%%%%%%%%%%%%%%%%%%%%%%
% 
% Analytical model enables quick scan of parameter space to find  $q$ and $\fscC$ values that correspond to waveguide with particularly desirable features, a CIP in our case. Analytical dispersion is predicted using Mathiue Exponent $\kscale=\kscale\left(\fsc,\fscC,q\right)$; this approximation works well around the centre of the waveguide and predicts field shapes quite well, especially in case of shallow corrugations, i.e. $q \ll 1$ and $L_0 \ll L_z$. We choose corrugation parameter $q$ to be $0.1$, and find that analytical model predicts that CIP occurs for $\fscC=1.255$, i.e.
%\vspace*{-0.25em}

Using numerical simulations is an alternative way of calculating dispersion relation of the waveguide, as well as checking the validity of the analytical predictions. We employ commercial numerical simulation software package CST that solves discretised Maxwell integral equations on a tetrahedral mesh. 
We simulate corrugated waveguiding structure described above which is one or several periods long, depending on the particular aim of the simulation. We aim to numerically find the CIP theoretically predicted in the previous section, and to numerically confirm  behaviour of the EM modes suitable for the particle interaction.
We start by simulating ten period (10 $L_z$) long waveguide using an eigenmode solver. We set boundary conditions as perfect electric conductor (PEC) on the walls of the waveguide, and as quasiperiodic conditions at the entrance and exit of the waveguide, i.e. 
\vspace*{-0.25em}
\begin{align}
\VEft_\parallel |_{\textup{Bdd}} = 0, \qquad \VEft\!\left(z=0\right)=e^{i \theta}\VEft\!\left(z=10L_z\right). 
\label{QuasiPeriodic_Num_Disp}
\end{align}

\begin{figure}[htbp]
\centering\includegraphics[width=0.74\textwidth]{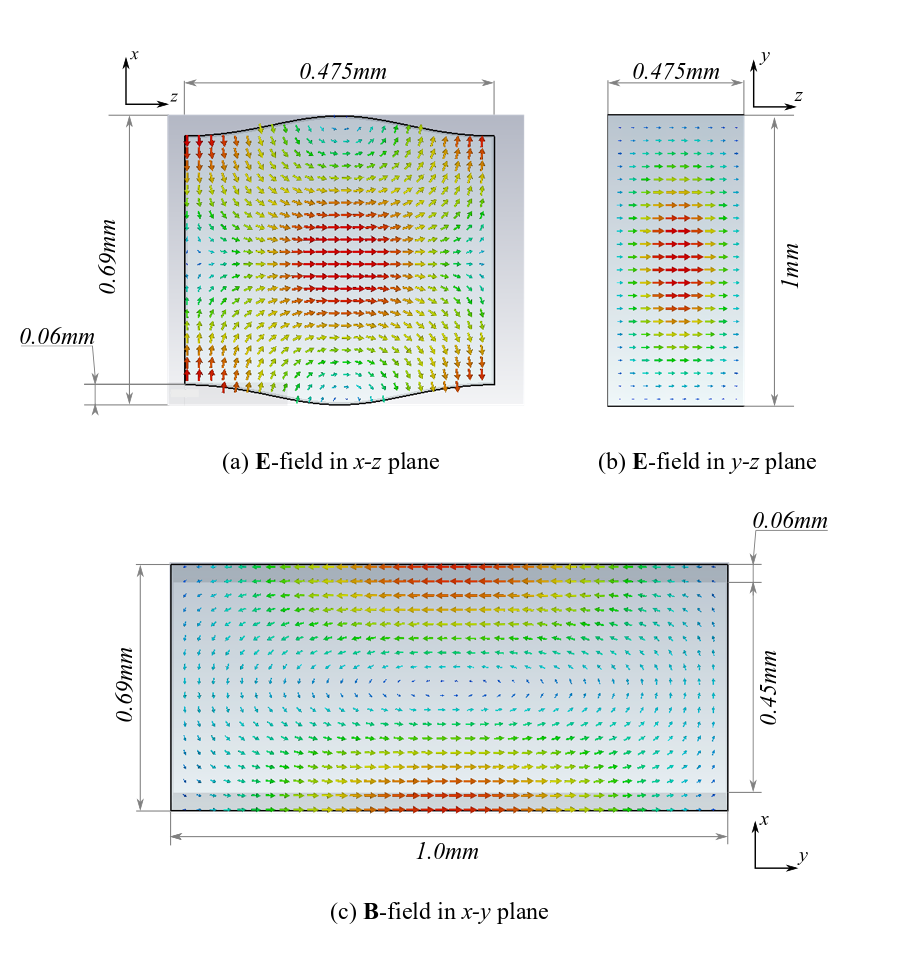} 
\caption{Example of the simulated electric and magnetic field for the
  corrugated structure. We see that \textbf{E}-field is strongest at the centre of the waveguide; it is also collinear with $z$-axis.}
\label{fig_3PFPF}
\end{figure}

Firstly, in simulation results, we find field pattern that correpsonds to TM mode, -- \textbf{E}-field concentrated in the centre of the waveguide and collinear with $z$ axis, and \textbf{B}-field confined in the cross-section of the waveguide. An example of such field pattern, namely crossections of $\textbf{E}$-field in $x-z$ and $y-z$ planes and $\textbf{B}$-field in $x-y$ plane, found in CST simulations is given on Figure \ref{fig_3PFPF}. Geometrical parameters of the waveguide used in the simulations correspond to numerical CIP (parameter values are $\Lz=0.475 \mmeters$, $\Ly=1 \mmeters$, $\Lnul=0.5 \mmeters$ and $q=0.1$, and synchronous particle velocity in this case is 0.46).

%%%%%%%%%%%%%%%%%%%%%%%%%%%%%%%%%%%%%%%%%%%%%%%%%%%%%%%%%%%%%%%%%%%%%%
%              3PROJECTIONS FIELD PATTERNS FIGURE 
%%%%%%%%%%%%%%%%%%%%%%%%%%%%%%%%%%%%%%%%%%%%%%%%%%%%%%%%%%%%%%%%%%%%%%

%%%%%%%%%%%%%%%%%%%%%%%%%%%%%%%%%%%%%%%%%%%%%%%%%%%%%%%%%%%%%%%%%%%%%%
%
%%%%%%%%%%%%%%%%%%%%%%%%%%%%%%%%%%%%%%%%%%%%%%%%%%%%%%%%%%%%%%%%%%%%%%

\begin{figure}[htbp]
\begin{tikzpicture}
\draw(0,0) node[above right=-4pt] {\includegraphics[width=1\textwidth]{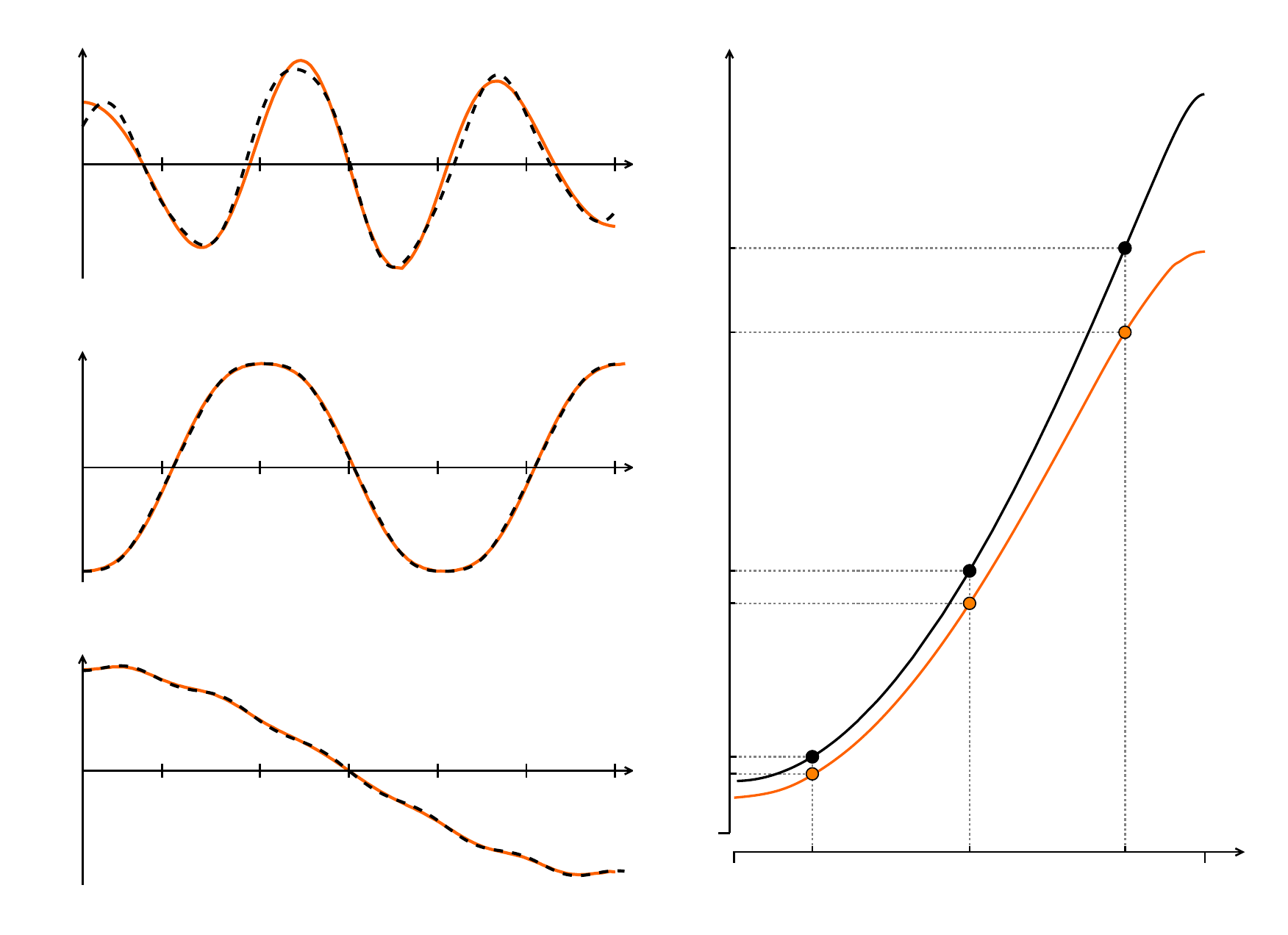}} ;
\draw(0.59,8.03) node{$\VEft$} ;
\draw(0.59,4.91) node{$\VEft$} ;
\draw(0.59,1.72) node{$\VEft$} ;
% \draw(0.59,8) node{$E_z$} ;
% \draw(0.59,4.8) node{$E_z$} ;
% \draw(0.59,1.7) node{$E_z$} ;
\draw(6.545,8.19) node{$\zeta$} ;
\draw(6.545,5.06) node{$\zeta$} ;
\draw(6.545,1.92) node{$\zeta$} ;
% \draw(6.325,4.64) node{$z$} ;
% \draw(6.325,1.48) node{$z$} ;
\draw(4.1,9) node{$\kscale=2.83$} ;
\draw(4.4,6) node{$\kscale=2.5$} ;
\draw(3,2.7) node{$\kscale=2.17$} ;
% RIGHT SUBFIGURE
\draw(7.86,9) node{\large $\fsc$} ;
\draw(12.79,1.15) node{\large $\kscale$} ;
\draw(7.62,0.55) node{\small 2.0} ;
\draw(8.50,0.55) node{\small 2.17} ;
\draw(10.05,0.55) node{\small 2.5} ;
\draw(11.74,0.55) node{\small 2.83} ;
\draw(12.47,0.55) node{\small 3.0} ;
%\draw(7.215,1.055) node{\small 1.2} ;
\draw(7.205,1.84) node{\small 1.07} ;
\draw(7.205,3.412) node{\small 1.15} ;
\draw(7.285,6.225) node{\small 1.3} ;
\draw(7.205,7.09) node{\small 1.34} ;
% SUBPICTURES LETTERS
\draw(6.2,9) node{\normalsize (a)} ;
\draw(6.2,6.09) node{\normalsize (b)} ;
\draw(6.2,2.75) node{\normalsize (c)} ;
\draw(12.7,9) node{\normalsize (d)} ;
%LEFT X-axis tick labels
\draw(6.37,7.75) node{\small $6\pi$} ;
\draw(6.4,4.59) node{\small $6\pi$} ;
\draw(6.4,1.44) node{\small $6\pi$} ;
\draw(1.7,8.14) node{\small $\pi$} ;
\draw(1.7,5.04) node{\small $\pi$} ;
\draw(1.7,1.89) node{\small $\pi$} ;
\draw(4.5,8.16) node{\small $4\pi$} ;
\draw(4.5,5.04) node{\small $4\pi$} ;
\draw(4.5,1.89) node{\small $4\pi$} ;
\end{tikzpicture}
\caption{Longitudinal field profiles  are shown along the centre of the waveguide $E_z\left(x\!=\!0,y\!=\!0,z\right)$ on the subfigures (a)-(c). On the left longitudinal component of the electric field along the centre of the waveguide for the 6 periods long structure at 3 different wavenumber values. Numerical (orange) versus predicted analytical (black dashed).  Three pairs of  corresponding points on the numerical (orange) and analytical(black solid)  dispersion curves are shown on the right (d). We observe good agreement between the field patterns for the most of the zone, apart from the region of higher frequencies as we start approaching the bandgap.}
\label{fig_Siml_NvP_E}
\end{figure}
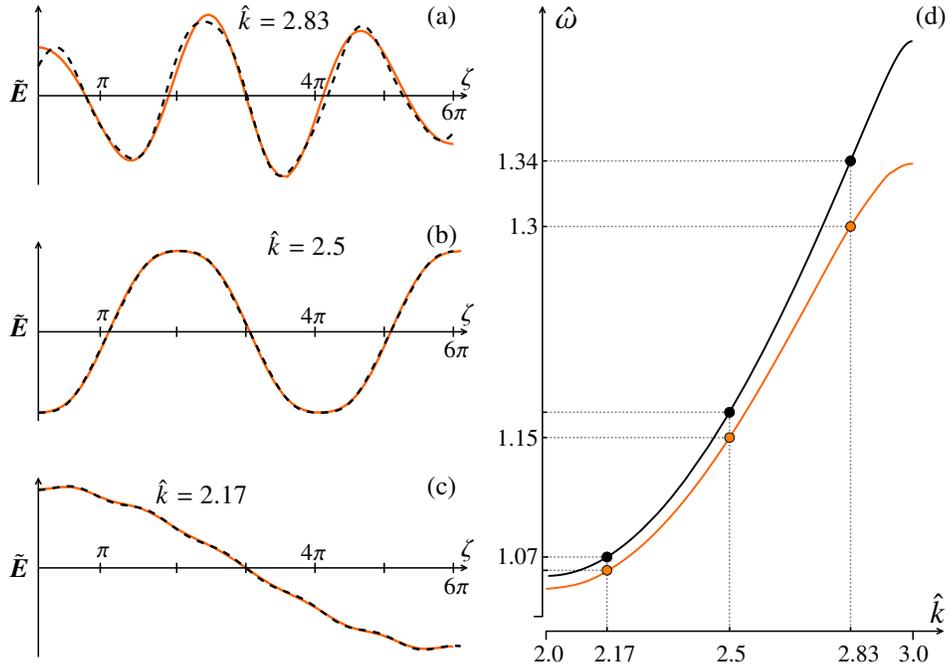

We proceed to compare analytical and numerical predictions for dispersion relations and field profiles in the centre of the waveguide.  Analytical dispersion (shown as a black line on the figure \ref{fig_Siml_NvP_E}(d)) is calculated as a Mathiue exponent, while numerical dispersion (orange line) $\omega (k(\theta)) $ is found by varying phase $\theta$ in quasiperiodic boundary conditions (\ref{QuasiPeriodic_Num_Disp}) and using wavenumber-phase relation $\kscale=2+\theta/\left(10\pi\right)$.
The third zone of dispersion relation, as predicted by analytical calculation and numerical simulations is given in the Figure \ref{fig_Siml_NvP_E}(d). Line of synchronous particle, travelling at  $0.46$ speed of light, is shown in blue. We note good agreement between numerical and analytical dispersion, with an exception of a region close to the bandgap.

To compare field profiles in the centre of the waveguide obtained in numerical calculations and those predicted by analytical theory we need set a definite phase of the EM mode in the waveguide. To this end, we consider a 6-period long corrugated structure with PEC conditions at the entrance and exit of the waveguide, i.e. find standing modes of the resonator, rather than a waveguide. This enables us to find shapes of the longitudinal electric field $E_z\left(z\right)$ in the centre of the waveguide  for several points in the third zone of dispersion, with wavenumbers $\kscale=2.17$, $\kscale=2.5$  and $\kscale=2.83$. Results, calculated for analytical CIP parameters ($L_z=0.475 \mmeters$, $L_y=1 \mmeters$, $\Lnul=1 \mmeters$, and $q=0.1$),  are shown in Figure \ref{fig_Siml_NvP_E} (a)-(c). We observe very good agreement for  $\kscale=2.17$ and $\kscale=2.5$, and acceptable agreement in the field shape even for the point $\kscale=2.83$, where the discrepancy between analytical and numerical dispersion curves becomes greater, as we are nearing the bandgap.

%%%%%%%%%%%%%%%%%%%%%%%%%%%%%%%%%%%%%%%%%%%%%%%%%%%%%%%%%%%%%%%%%%%%%%
%            DISPERSION  aCIP and FIELDS
%%%%%%%%%%%%%%%%%%%%%%%%%%%%%%%%%%%%%%%%%%%%%%%%%%%%%%%%%%%%%%%%%%%%%%

%%%%%%%%%%%%%%%%%%%%%%%%%%%%%%%%%%%%%%%%%%%%%%%%%%%%%%%%%%%%%%%%%%%%%%
%
%%%%%%%%%%%%%%%%%%%%%%%%%%%%%%%%%%%%%%%%%%%%%%%%%%%%%%%%%%%%%%%%%%%%%%

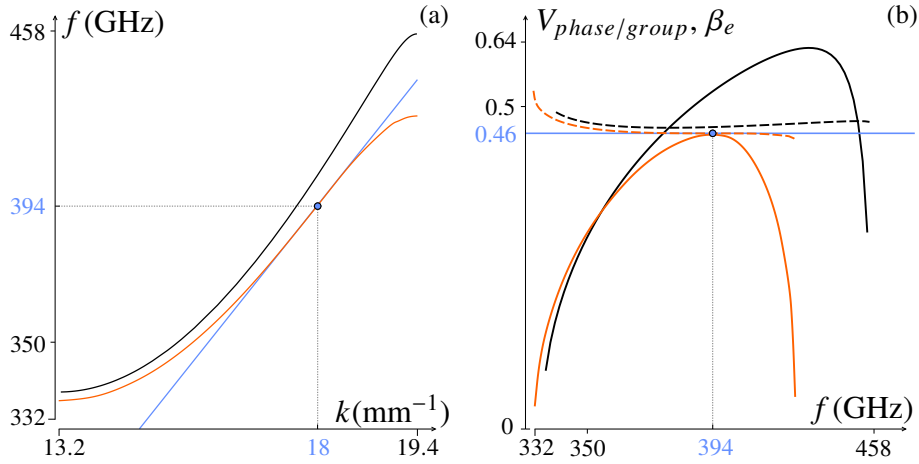
\begin{figure}[htbp]
\begin{tikzpicture}[scale=1]
\draw (0,0) node[above right=0cm]{
\centering\includegraphics[width=0.95\textwidth]{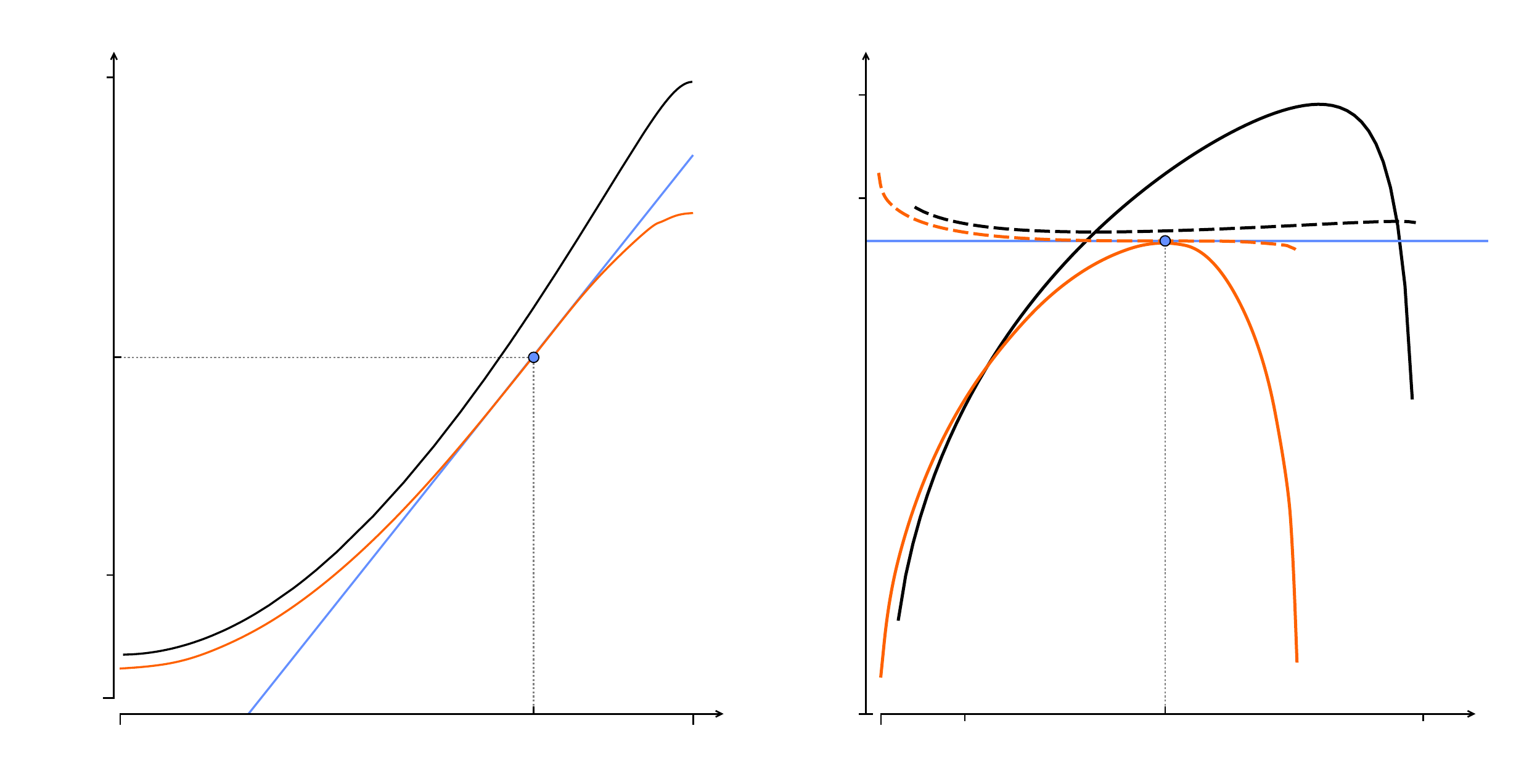} 
};
%1ST SUBfigure
\draw(1.84,6.09) node{\large $f\!$ (GHz)} ;
\draw(5.55,0.95) node{\large $k$(mm$^{-1}$)} ;
\draw (0.71,5.99) node {\small $458$} ;
\draw (0.71,3.67) node {\small \textcolor{blueColPal}{$394$}} ;
\draw (0.71,1.84) node {\small $350$} ;
\draw (0.71,0.87) node {\small $332$} ;
\draw (1.19,0.46) node {\small $13.2$} ;
\draw (4.57,0.46) node {\small \textcolor{blueColPal}{$18$}} ;
\draw (5.89,0.46) node {\small $19.4$} ;
%2ND SUBfigure
\draw(11.79,0.95) node{\large $f\!$ (GHz)} ;
\draw(8.8,6.09) node{\large $V_{phase/group}$, $\beta_e$} ;
\draw(6.93,5.88) node{\small 0.64} ;
\draw(6.99,4.99) node{\small 0.5} ;
\draw(6.93,4.63) node{\small \textcolor{blueColPal}{0.46}} ;
\draw(7.08,0.75) node{\small 0} ;
\draw (11.95,0.46) node {\small $458$} ;
\draw (9.85,0.46) node {\small \textcolor{blueColPal}{$394$}} ;
\draw (8.15,0.46) node {\small $350$} ;
\draw (7.47,0.46) node {\small $332$} ;
% SUBPICTURES LETTERS
\draw(6.11,6.19) node{\normalsize (a)} ;
\draw(12.33,6.19) node{\normalsize (b)} ;
\end{tikzpicture}
\centering
\caption{Dispersion relations, numerical (orange) and analytical (black),  for numerical CIP parameters are shown on subfigure (a). On subfigure (b) numerical phase (dashed orange) and group (solid orange) velocities are shown, and synchronous particle line (solid blue) are shown toghether with analytical predictions ( paremeter values $q=0.1$, $\fscC=1.06$) of phase and group velocities (dashed and solid black lines correspondingly).  Numerical CIP itself is marked as a blue point.}
\label{fig_n_VCIP}
\end{figure}

Finally, to check how well the CIP conditions (\ref{CIP_explicitly}) are met, we calculate phase and group velocities from analytical and numerical dispersions,-- shown in Figure \ref{fig_n_VCIP}(a). On Fig. \ref{fig_n_VCIP}(b) we show how numerical phase velocity (orange dashed line), numerical group velocity (orange solid line), and particle velocity (blue line) are matched in the CIP. Analytical results are calculated for the parameters that correspond to numerical CIP($q=0.1$, $\fscC=1.06$).  It is evident that analytical phase velocity (black dashed line) is  close to the numerical results, whereas there is a considerable discrepancy  in the group velocity.

We observe that exact CIP is a strong condition on the shape of the dispersion curve. Achieving exact CIP for numerical 3D dispersion we had to change value of $\Lnul$ from $0.41 \mmeters$ to $0.5 \mmeters$ comapred to analytical CIP (i.e. changing $\fscC$ from $1.255$ to $1.06$). This results in CIP synchronous particle velocities and frequencies being different, $\beta_e=0.46$ and 394 GHz for numerical CIP, and $\beta_e=0.53$ and 474 GHZ for analytical CIP.
Both numerical and anlytical estimations relying on different assumptions and  are approximations of the real system.

%  To achieve exact numerical CIP geometric parametes of the structure must be varied. As $L_z$, i.e. length of the corrugation period, determines the mode wavenumber and overall scaling we keep it constant. Changes in $L_y$, width of the waveguide, have small effect on the value of the band cut-off frequency $\fscC$. Thus, we are left with two parameters to vary, $q$ and $\Lnul$, both affecting the depth of corrugation. It turns out that varying $\Lnul$ requires smaller relative change. To achieve exact CIP for numerical 3D dispersion we have to change value of $\Lnul$ from $0.41 \mmeters$ to $0.5 \mmeters$ (i.e. changing $\fscC$ from $1.255$ to $1.06$).

%%%%%%%%%%%%%%%%%%%%%%%%%%%%%%%%%%%%%%%%%%%%%%%%%%%%%%%%%%%%%%%%%%%%%%
%            N--VCIP
%%%%%%%%%%%%%%%%%%%%%%%%%%%%%%%%%%%%%%%%%%%%%%%%%%%%%%%%%%%%%%%%%%%%%%

\section{Waveguide Mediated Particle Wave Interaction} %Energy interaction between wave and electron beam}
\label{ch_Force}

In this section, we employ two approaches to examine the particle-wave interaction as described in section \ref{ch_Sol} above. The first is based on a single particle model that remains in phase with the EM wave. This demonstrates the relationship between the dispersion relation and the velocity of the particle. The second approach is based on a perturbation technique, commonly referred to as Madey's theory \cite{yan}. This approach combines a perturbation expansion of the Lorentz equation  and averaging over the initial phases to study energy exchange between beam and wave.

\begin{figure}[tb]
\centering
\begin{tikzpicture}[scale=0.95]
\draw (0,0) node[above right=-.5cm]{
\includegraphics[width=5.85cm]{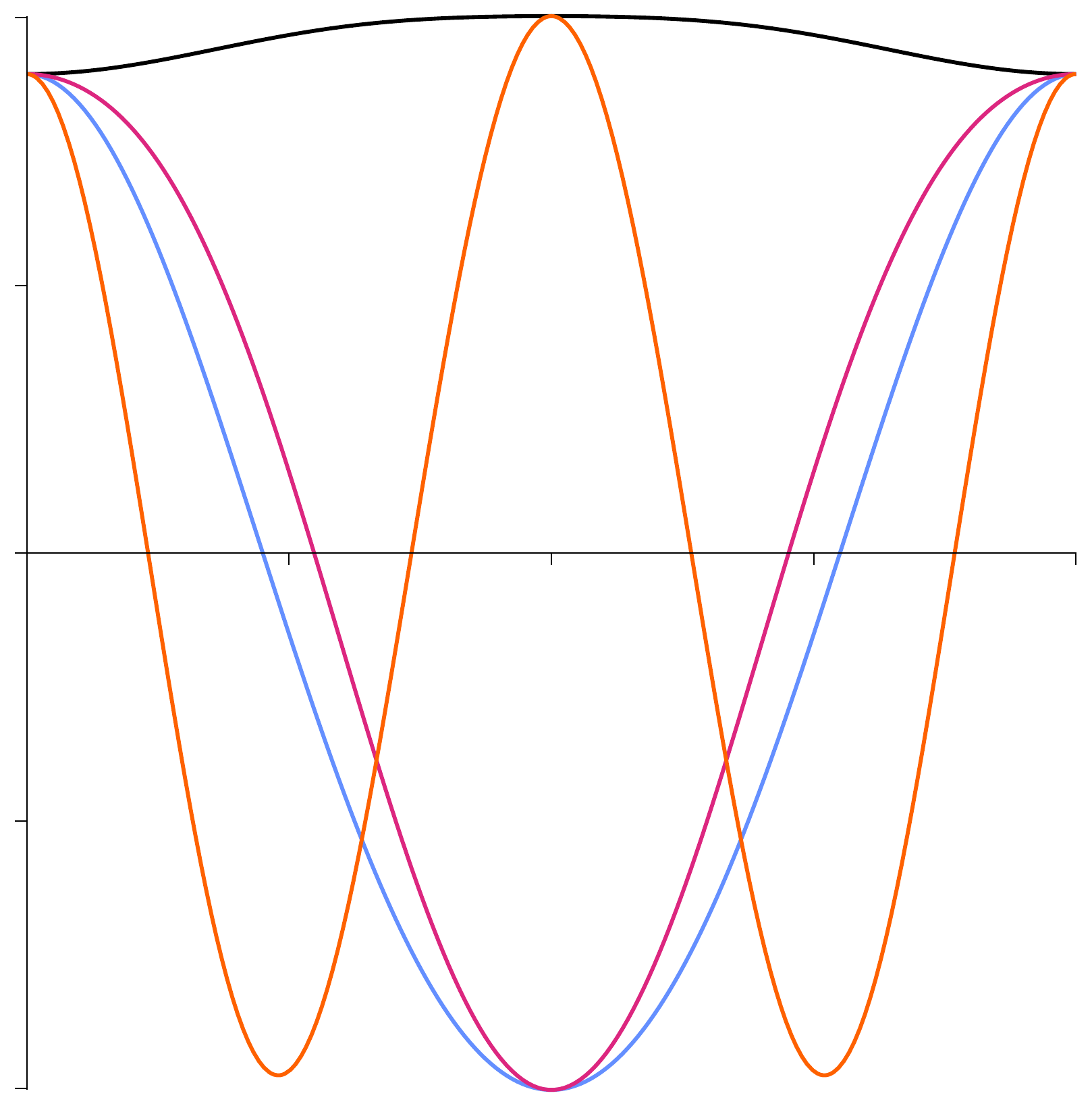}} ;
\draw (3,6.0) node {\small $1^{\text{st}}$ zone, $\kscale=0.857$} ;
\draw (0.94,2.55) node[rotate=-71.5,blueColPal] 
  {\small $2^{\text{nd}}$ zone, $\kscale=1.143$} ;
\draw (1.2485,3.8) node[rotate=-67,magentaColPal] 
  {\small $3^{\text{rd}}$ zone, $\kscale=2.857$} ;
\draw (3.47,4.2) node[rotate=-77,orangeColPal] 
  {\small $4^{\text{th}}$ zone, $\kscale=3.143$} ;

\begin{scope}[shift={(-0.05,2.85)},yscale=3]
\draw (-0.2,-1.03) node[left] {-1.0} ;
\draw (-0.2,-0.53) node[left] {-0.5} ;
\draw (-0.2,-0.035) node[left] {0} ;
\draw (-0.2,0.47) node[left] {0.5} ;
\draw (-0.2,0.975) node[left] {1.0} ;
\end{scope}
\draw (-1.25,4.9) node[left,rotate={90}] {\large force\ (eV\,m$^{-1}$)} ;

\begin{scope}[shift={(0,2.8)}]
%\draw (0,0) node [below] {0} ;
%\draw (1.5,0) node [below] {$\tfrac14\pi$} ;
\draw (2.7,-0.1) node [below] {$\tfrac\pi2$} ;
%\draw (4.5,0) node [below] {$\tfrac34\pi$} ;
\draw (5.65,-0.1) node [below] {$\pi$} ;
\draw (5.5,.52) node [below] {\large $\zeta$} ;
\end{scope}
\end{tikzpicture}
\caption{The force felt by an in-phase electron in the first four zones. The strength of the electric field is 1Vm$^{-1}$ at the peak part of the waveguide. The zone is determined by the speed of the particle. The first and second zones are inaccessible since they corresponds to superluminal particles.
 Here $q=0.1$, $\fscC=1.255$ and  $\fsc=1.513$.
First  zone (black), $\kscale=0.857$,
  \mbox{$\beta_e=1.766$}. Total Force \mbox{$\forcePeriod=4.81$}.
  Observe that the $\text{force}>0$ for all
  $\zeta$.  
Second  zone (blue), $\kscale=1.143$, $\beta_e=1.324$,
$\forcePeriod=-0.516$. 
Third  zone (red). $\kscale=2.857$, $\beta_e=0.530$,
 $\forcePeriod=0.244$. This is at the CIP.
Fouth zone (green), $\kscale=3.142$, $\beta_e=0.481$, $\forcePeriod=-0.043$.}
\label{fig_Mat_Force}
\end{figure}
%%%%%%%%%%%%%%%%%%%%%%%%%%%%%
%
%%%%%%%%%%%%%%%%%%%%%%%%%%%%%

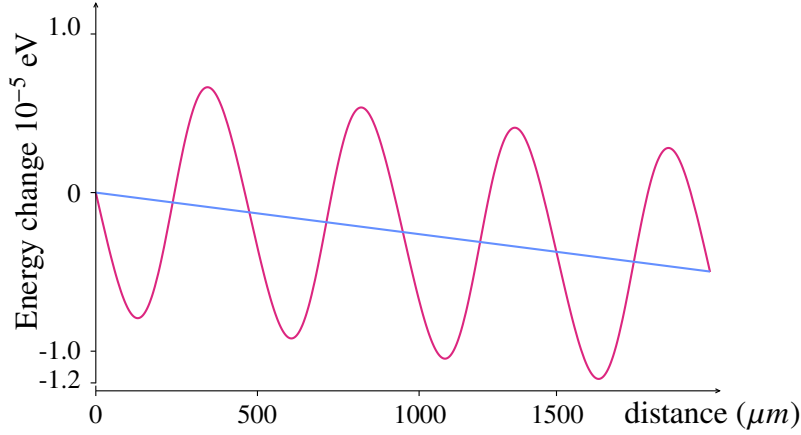
\begin{figure}[tb]
\centering
\begin{tikzpicture}[scale=0.95]
\draw (0,0) node[above right=-.5cm]{
\includegraphics[width=0.7\textwidth]{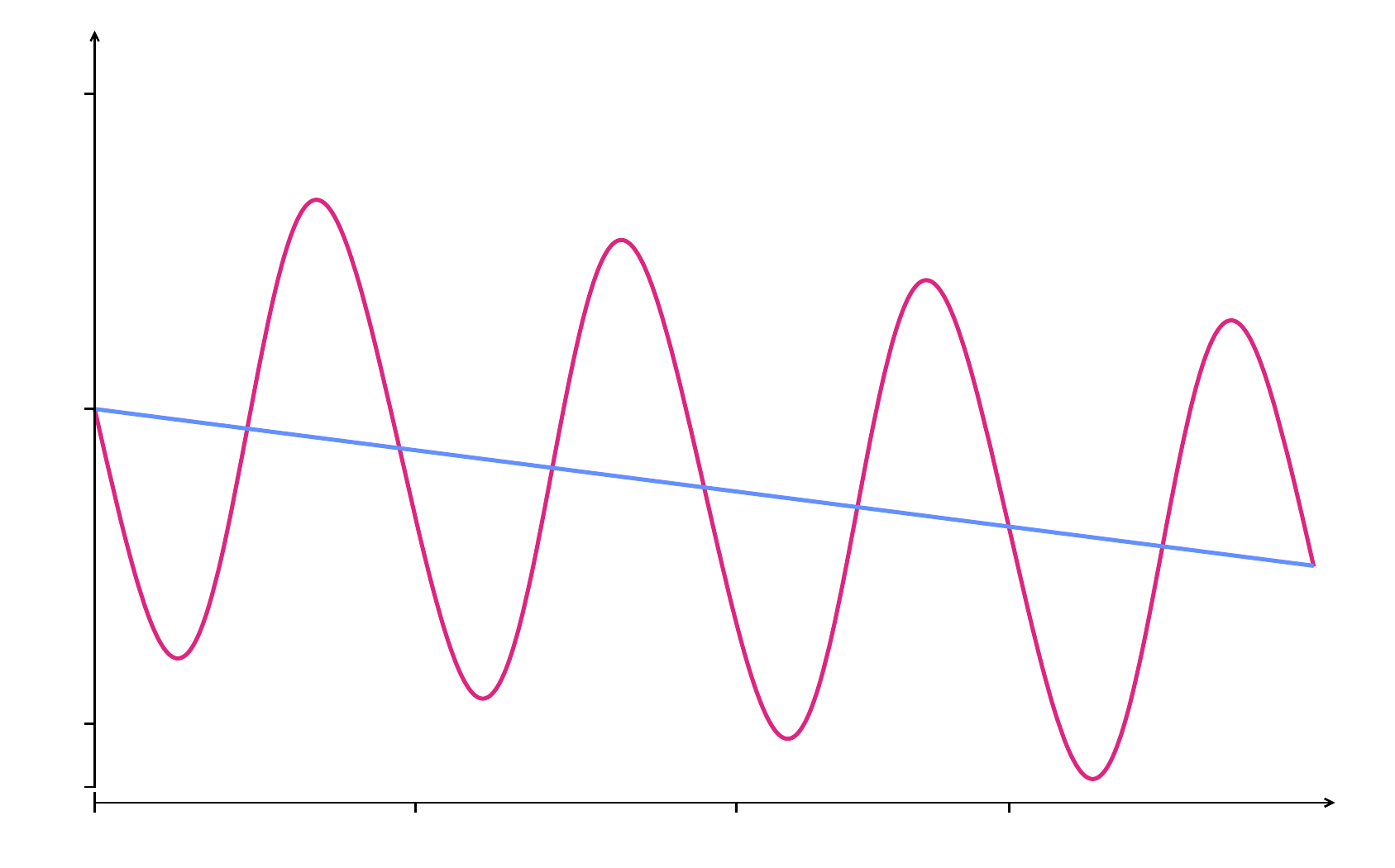}} ;

\begin{scope}[shift={(-0.05,2.85)},yscale=3]
\draw (0.24,-0.875) node[left] {-1.2} ;
\draw (0.24,-0.74) node[left] {-1.0} ;
\draw (0.24,-0.005) node[left] {0} ;
\draw (0.24,0.775) node[left] {1.0} ;
\end{scope}

%%%%%%%%%%%%%%%%%%%%%%%%%%%%%

\begin{scope}[shift={(0,2.8)}]
\draw (-0.7,0.2) node[rotate={90}] {\large Energy change $10^{-5}$ eV} ;
%\draw (-0.9,1) node[rotate={90}] {$10^{-5}$ eV} ;
\draw (0.27,-2.8) node [below] {0} ;
\draw (2.53,-2.8) node [below] {500} ;
\draw (4.805,-2.8) node [below] {1000} ;
\draw (6.725,-2.8) node [below] {1500} ;
\draw (8.9,-2.75) node [below] {\large distance ($\mu m$)} ;
\end{scope}
\end{tikzpicture}
\caption{Change in kinetic energy of a single electron, from 92KeV(i.e. synchronous beam, $\beta_0=0.53$, third zone in figure \ref{fig_Mat_Force}), in an RF field of 1Vm$^{-1}$, 
over four periods in the waveguide. Here the waveguide is at the {analytical} CIP for $q=0.1$, and {geometric parameters $L_z=0.475 \mmeters$, $L_y=1 \mmeters$, $\Lnul=1 \mmeters$}. The blue line is the average energy loss.}
\label{fig_IntForce}
\end{figure}

% \begin{figure}[htbp]
% \centering
% \begin{tikzpicture}[scale=0.9]
% \draw (0,0) node[above right=-.5cm]{
% \includegraphics[width=5.85cm,height=2.9cm]{./FiguresOE/fig8_q1}} ;

% \begin{scope}[shift={(-.05,1.8)},yscale=1.94]
% \draw (0,-1) node[left] {-1.0} ;
% %$\draw (0,-2) node[left] {-2.0} ;
% %\draw (0,-3) node[left] {-3.0} ;
% \draw (0,0) node[left] {0.0} ;
% %\draw (0,1) node[left] {1.0} ;
% %\draw (0,2) node[left] {2.0} ;
% %\draw (0,3) node[left] {3.0} ;
% \draw [color=blue,very thick] (-0.01,0.01) -- (3.75*1.6,-.38) ;
% \end{scope}
% \draw (-1.3,1) node[rotate={90}] {Energy change} ;
% \draw (-0.9,1) node[rotate={90}] {$10^{-5}$ eV} ;

% \begin{scope}[shift={(-0.0,-0.1)},xscale=1.6]
% \draw (0,0) node [below] {0} ;
% \draw (1,0) node [below] {$500$} ;
% \draw (2,0) node [below] {$1000$} ;
% \draw (3,0) node [below] {$1500$} ;
% \draw (3,-.5) node [below] {distance ($\mu m$)} ;
% \end{scope}
% \end{tikzpicture}
% \caption{
% Change in kinetic energy of a single electron, from 92KeV(i.e. synchronous beam, $\beta_0=0.53$, third zone in figure \ref{fig_Mat_Force}), in an RF field of 1Vm$^{-1}$, 
% over four periods in the waveguide. Here the waveguide is at the CIP for $q=0.1$, with the parameters given in Fig. \ref{fig_Mat_Force} and scaled with $L_z$, $L_0$ and $L_y$ from figure \ref{fig_Mat_Force}. The blue line is the average energy loss.}
% \label{fig_IntForce}
% \end{figure}

As suggested in the previous section $E_z$ field shape obtained from analytical model closely resembles field shape produced in numerical  simulations, and in this section we will use analytical field expressions to estimate wave particle interaction in the structure.

To examine the interaction between charged particles and EM waves in our structure we start by considering a single particle (charge $Q$) moving along a trajectory in the
centre of the waveguide.
In general the on-axis ($(x=0,\, y=0)$ electric field $E(t,z)$ within the waveguide at time $t$ and position $z$ is given by inverse fourier transform 
\[
E(t,z) = \frac{1}{2\pi}\int   \VEft(\omega,z) e^{-i\omega t}d\omega
\]
A particle that enters the waveguide at time $t_o$, the particle will be at position $z_p$ at time $t=t_o + z/c\beta_e \equiv t_p$. Therefore the field experienced by the particle as it travels through the waveguide is
\[
E(t_p,z_p) = \frac{1}{2\pi}\int   \VEft(\omega,z_p) e^{-i\omega (t_0 + z/c\beta_e)} d\omega
\]
Substituting the the on-axis field from equation~\ref{Sol_E}, 
\[
\VEft(\omega,z_p) = B_0 (\kapx^2+\lamy^2)\int e^{i k z} \MatPF_{a,q,\kscale} \big(\pi\,\Lz^{-1}\,z\big),
\]
the field experienced becomes
\begin{equation}
\begin{aligned}
 E(t_p ,z_p)  & = \frac{1}{2\pi}B_0 (\kapx^2+\lamy^2)\int e^{i k z_p} \MatPF_{a,q,\kscale} \big(\pi\,\Lz^{-1}\,z_p\big) e^{-i\omega (t_0 + z_p/c\beta_e) }\rho(\omega)d\omega  \\
%& = \frac{1}{2\pi}B_0 (\kapx^2+\lamy^2)\int \MatPF_{a,q,\kscale} \big(\pi\,\Lz^{-1}\,z\big) e^{i\omega t_0 }d\omega  \\
\end{aligned}
\end{equation}
where we have introduced the spectral density of the waveguide field, $\rho(\omega)$. For a monochromatic field with $\rho(\omega) = \delta(\omega-\omega_s) + \delta(\omega-\omega_s)$ at the phase-matched frequency $\omega_s$ we obtain the effective field to be
\begin{equation}
\begin{aligned}
E(t_p ,z_p)  & = \frac{1}{\pi}B_0 (\kapx^2+\lamy^2) \Re \left[\MatPF_{a,q,\kscale} \Big(\frac{\pi\,z_p}{\Lz} \Big) \right] \cos(\omega_s t_0)    \\
& =   \pi B_0 \Lz^{-2}
\left(\fscC^2  + 2 \,q\,\cos(2\pi\,\Lz^{-1}\,z_p)\right) \Re \left[\MatPF_{a,q,\kscale} \Big(\frac{\pi\,z_p}{\Lz} \Big) \right]\cos(\omega_s t_0)    
\end{aligned}
\end{equation}
where we have used the velocity phase-matched condition 
 $k = \omega_s /c\beta_e$, and the expansion of $\kapx$ from equations~(\ref{Sol_lambda_z}) and (\ref{Match_Lx_L0}).

The energy gain of a particle traveling on-axis over one structure period is therefore
\begin{align}
U &= Q\int_0^{L_z}E(t_p,z_p) \,dz_p  \nonumber \\
&= \frac{ Q \,B_0 \pi\,\cos(\omega_s t_0) }{L_z^2} \int_0^{L_z}  \left(\fscC^2  + 2 \,q\,\cos\Big(\frac{2\pi\,z_p}{\Lz} \Big) \right) \Re \left[\MatPF_{a,q,\kscale} \Big(\frac{\pi\,z_p}{\Lz} \Big)  \right] \, dz_p 
\end{align}

\color{black}

%%%%%%%%%%%%%%%%%%%%%%%%%%%%%%%%%%%%%

From figure \ref{fig_Mat_Force} we see the energy depends on
which zone the interaction occurs in, determined by
the velocity of the particle. In all cases $\forcePeriod$ is highest when
we are in the first zone, since there is no counter
force. However, we show in Supplemental Document, in the section SD3 that this is
impossible, as it requires
superluminal particle velocities.
The total interaction of the backward
wave (second zone) is  comparable to the forward wave
(third zone). In this case the backward wave would
also need superluminal particle velocities. By comparison the forward wave is
subluminal, and has a very good interaction between wave and particles, equal
to $12\%$ of the first zone interaction. Over several periods the energy of the particle shows a net decreases as seen in figure \ref{fig_IntForce}, and a particle synchronised with the opposite phase would exhibit a net increase in energy.

%%%%%%%%%%%%%%%%%%%%%%%%%%%%%%%%%%%%%%%%%%%%%%%%%%%%%%%%%%%%%%%%%%%%%%
%\subsection{Madey's theroem}

 We extend this examination %of %waveguide mediated  charged particle EM wave interaction by energy transfer 
 by considering the energy transfer between a charged particle beam and the EM wave using Madeys theorem\cite{MadeyOrig,yan}, which  %Madeys theorem 
 has been used %to  examine energy exchange between particle beams and EM wave 
 in systems, from free electron lasers\cite{yan,xfel}, to conventional traveling wavetubes\cite{yan}, to metamaterial vacuum electronic devices\cite{sev1}.
  Madeys theorem relates the phase averaged energy spread to the phase averaged energy
change experienced by a charged particle as it propagates through a system. The theory uses the first two terms of a perturbation expansion of the Lorentz force
based on a single charged particle analysis where (ignoring space charge) a uniformly distributed charged particle beam is injected into the system.
The theorem draws links between the spontaneous emission of photons by a single electron passing through the structure to stimulated emission of photons. 
The
first perturbation energy term of the Lorentz equation
$\gamma_1$ is taken at entry and exit from the structure, the difference $\Delta \gamma_1$ is averaged over the phase of the EM
field to yield equation (\ref{eq:m1}). This change in energy $\Delta \gamma_1$ relates to the classical spontaneous power spectrum from
Maxwell's equations from the beam\cite{yan},
\begin{equation}
\left< \Delta \gamma^{2}_{1}\right>= \left< \left( \int\!\!\!dz \frac{-Q E_z \beta_0}{m_0 c} \right)^2 \right>.
\label{eq:m1}
\end{equation}

The second order %perturbation 
term relates the energy change in the beam due to a stimulated emission response, a consequence of a generalized framework in Hamiltonian mechanics \cite{LuchiniMotzFEL}, given by,
\begin{equation}
\left< \Delta \gamma_{2}\right>= \frac{1}{2} \frac{d}{d\gamma} \left< \Delta \gamma^{2}_{1}\right>.
\label{eq:m2}
\end{equation}
In a simplistic model we can consider the electron beam as
 $N$ electrons entering the system every second, this enables us to
write the power change in the EM wave as,
\begin{equation}
\Delta P = -\frac{1}{2} \frac{d}{d\gamma} \left< \Delta \gamma^{2}_{1}\right> m_0 c^2 N.
\label{eq:m3}
\end{equation}

%{\color{red}****Two sentences on the form used by Sergey to plot graphs}

Using (\ref{eq:m3}) we calculate power change in the 10 period long waveguide (i.e. structure is 10 $L_z$ long). We assume that electron beam is in the centre of the waveguide and comprises of $N=10^4$ electrons; value EM field amplitude is taken as 1 V/m. We perform calculation for two energies of the beam: $\beta_0=0.53$ - which corresponds to the CIP as on Fig.2, and $\beta_0=0.55$ - which correspond to a single intersection point in the 3rd zone.

\begin{figure}
\centering
\begin{tikzpicture}
\draw(-.5,-.5) node[above right=6pt]{\includegraphics[width=1.\textwidth]{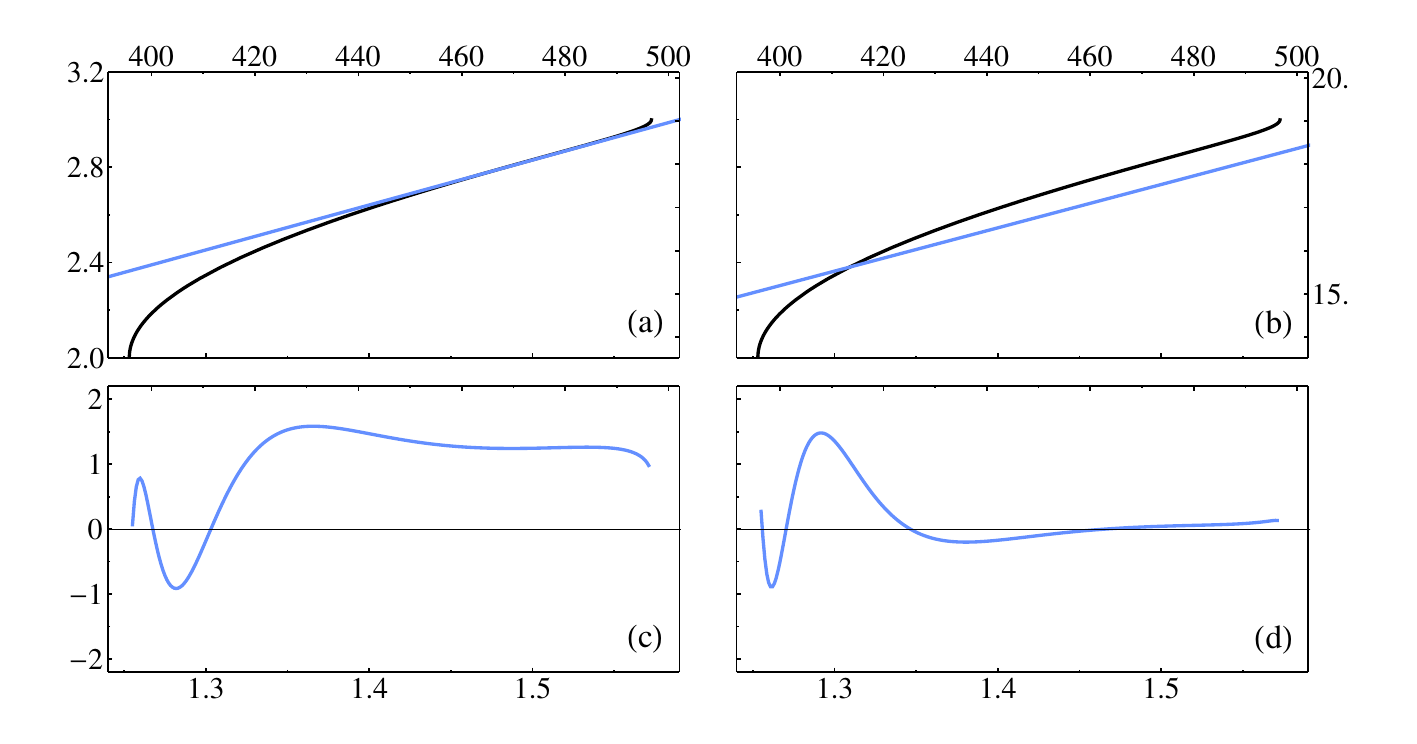}} ;
\draw(0.125,6.2) node{\large$\kscale$} ;
\draw(0.1,2.1) node[rotate=90]{$\Delta P$ (eV)} ;
\draw(5.9,0.23) node{\large$\fsc$} ;
\draw(11.9,0.23) node{\large$\fsc$} ;
\draw(12.7,5.1) node[rotate=90]{\large$k$ (mm$^{-1}$)} ;
\draw(3.6,6.7) node{$f$ (GHz)} ;
\draw(9.55,6.7) node{$f$ (GHz)} ;
\end{tikzpicture}
\caption{On subfigures (a) and (b)  beam lines (black) and third zone waveguide dispersion Relations (blue) are shown for coincident inflecton point and for a simple single intersection correspondingly. EM wave Power Change, calculated using Madey Theorem, is shown for a CIP (c) and CIP  (d).}
\label{MadeyTriplePoint_and_non-TriplePoint}
\end{figure}
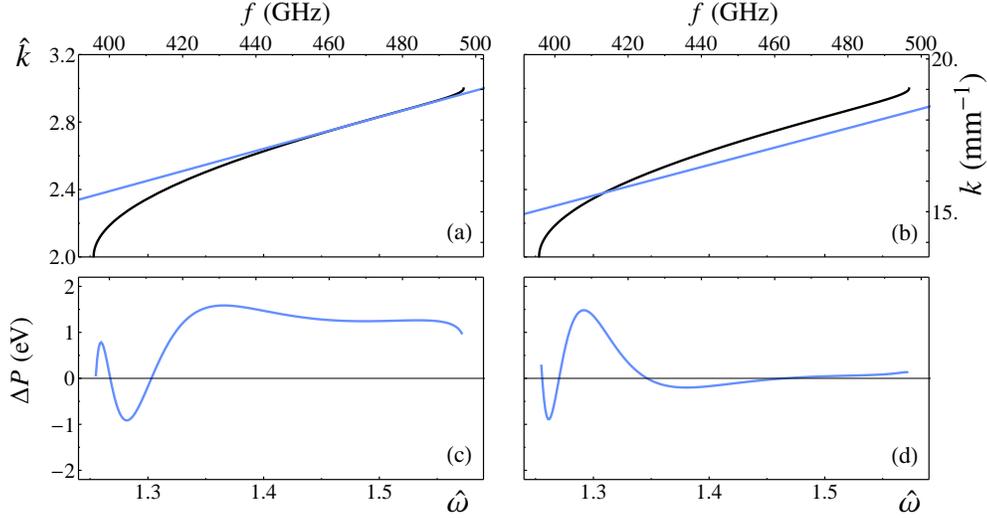

%{\color{red}****Discussion on Sergey  graphs}

Third zone dispersion relation and beam line for $\beta_0=0.53$, shown in \ref{MadeyTriplePoint_and_non-TriplePoint} (a), demonstrate considerable interval beam-wave interaction frequency interval. This interaction corresponds to the beam generating EM wave, as can be seen on \ref{MadeyTriplePoint_and_non-TriplePoint} (c), where, in the $\fsc$ frequency interval from approximately $1.32$ to $1.54$, positive sign of  $\Delta P$ corresponds to the power transferred from the electrons to EM wave. For the geometric dimensions of the waveguide used in the CST simulation corresponding frequency generation interval is 77GHz: from $419$ to $496$ GHz. Graph $\Delta P\left(\fsc\right)$  terminates as $\fsc$ frequency approaches band gaps.

The beam line for $\beta_0=0.55$ beam, shown in \ref{MadeyTriplePoint_and_non-TriplePoint} (b), crosses dispersion at a single point, without considerable interaction frequency interval. The EM power  $\Delta P$ (shown on \ref{MadeyTriplePoint_and_non-TriplePoint} (d)) is generated in a shorter frequency interval.

%{\color{red}***Noting regions of constant power change, inflections and acceleration/amplification }

%%%%%%%%%%%%%%%%%%%%%%%%%%%%%%%%%%%%%%%%%%%%%%%%%%%%%%%%%%%%%%%%%%%%%%

\section{Conclusion}
\label{ch_Concl}
We have shown it is possible to design corrugated waveguides where the phase velocity and group velocity of a wave coincide at a point of inflection in the phase velocity, which we refer to as the coincident inflection point (CIP). Using Madey's theorem we have shown the CIP creates an extended regime of interaction between the EM wave and a charged particle beam.
Even for the case where the CIP is close to the band gap (as seen in figure \ref{fig_Mat-Force_Disp})  the benefit of having a CIP is the frequency range of interaction extends over almost the whole band, as seen in figure \ref{MadeyTriplePoint_and_non-TriplePoint}.

The CIP has been found using a novel approach, based on an explicit analytic approximate solution  (\ref{Sol_B_flat}) and (\ref{Sol_E}) to Maxwell's equations, which involves Mathieu's equation. This allows us to estimate the geometrical parameters for waveguides with an engineered dispersion curve. 
We observe good agreement between the analytical dispersion curve and the CST numerical simulations. 

We show that waveguiding structure and synchronous particle beam at a CIP have potential for efficient broadband THz generation. We note again that waveguiding structures discussed in this manuscript have corrugations which mirror each other, figure \ref{fig_corr_waveguide}, and are distinct from sine waveguides in which crossection remains constant in shape and size and its centre undulates along the waveguide. 
In contrast, in our approach, the crossection of the waveguide varies and its centre remains constant.  We present a waveguide based traveling slow-wave structure that allows for an interaction between a charged particle beam with a velocity of 0.53c and a propagating EM wave with a longitudinal electric field.

This work can be extended to more general waveguides since the EM field given by equations (\ref{Sol_B_flat}) and (\ref{Sol_E}) are still good approximate a solution to Maxwell's equations even if the corrugation height is not given by equation (\ref{L_Approx_Form}) or equation (\ref{Match_Lx_L0}). Replacing equation (\ref{L_Approx_Form}) with a more general periodic function may offer the possibility of extending the CIP for higher velocities. One goal being to extend the maximum velocity, figure \ref{fig_Inf_v_q} to $c$, in order to exchange energy with ultra-relativistic particles. Furthermore, we can replace $\Lx(z)$ with any other functions as long as $\Lx'(z)$ and $\Lx''(z)$ are small, and hence we can apply these solutions for the EM field in the system that couples the corrugated waveguide to the external structure.  

Corrugated waveguides could also be used for the acceleration of charged particles. This work will be extended to investigate the use of our corrugated structures for particle acceleration, which  would only require a slight change in the synchronisation between wave and beam to flip this interaction from EM wave generation to particle acceleration \cite{sev1}.

%?? funnel of WG
%?? Talk about not a sin waveguide.

%Future work will focus on optimising parameters to maximise the interaction between wave and changed particle beams for both EM generation and amplification, and particle acceleration. As it would only require a slight change in the synchronisation between wave and beam to flip this interaction from EM wave generation to particle acceleration \cite{sev1}.

%\expandafter\show\the\font

\section{Backmatter}

\begin{backmatter}
\bmsection{Funding}
SSS, JG, SPJ and TB are grateful the support provided by STFC (the Cockcroft Institute ST/P002056/1 and
ST/V001612/1). JG is particularly grateful to the Peter Ratoff, director of Cockcroft (2014-2023), for supporting this research. 
RS gratefully acknowledges support from the AFRL Directed
Energy Chief Scientist Office and the EOARD, grant FA8655-20-1-7002.

%\bmsection{Acknowledgments}

\bmsection{Disclosures}
The authors declare no conflicts of interest.

\bmsection{Data availability} Data underlying the results presented in this paper are not publicly available at this time but may be obtained from the authors upon reasonable request.

\bmsection{Supplemental document}
See Supplemental Document for supporting content. 

\bmsection{Author contribution statement} JG suggested the original idea and provided the theoretical work. SJ and RS proposed applying the idea to THz generation. SJ proposed the idea of searching for the CIP. RS proposed the idea of using Madey's theorem, with SSS implementing it. The numerical simulations was principally undertaken by SSS based on initial work by TB. SSS lead the writing of the article, with help from all the authors. 

\end{backmatter}

%\section{Conclusion}
%After proofreading the manuscript, compress your .tex manuscript file and all figures (which should be in EPS or PDF format) in a ZIP, TAR or TAR-GZIP package. All files must be referenced at the root level (e.g., file \texttt{figure-1.eps}, not \texttt{/myfigs/figure-1.eps}). If there are supplementary materials, the associated files should not be included in your manuscript archive but be uploaded separately through the Prism interface.

%%%%%%%%%%%%%%%%%%%%%%% References %%%%%%%%%%%%%%%%%%%%%%%%%

%\bibliographystyle{abbrv}
\bibliography{referencesOE}

\begin{thebibliography}{10}
\newcommand{\enquote}[1]{``#1''}

\bibitem{Lemery2020}
F.~Lemery, T.~Vinatier, F.~Mayet, R.~A{\ss}mann, E.~Baynard, J.~Demailly,
  U.~Dorda, B.~Lucas, A.-K. Pandey, and M.~Pittman, \enquote{Highly scalable
  multicycle thz production with a homemade periodically poled macrocrystal,}
  {\protect\JournalTitle{Communications Physics}} \textbf{3}, 150 (2020).

\bibitem{Mosley2023}
C.~D.~W. Mosley, D.~S. Lake, D.~M. Graham, S.~P. Jamison, R.~B. Appleby,
  G.~Burt, and M.~T. Hibberd, \enquote{Large-area periodically-poled lithium
  niobate wafer stacks optimized for high-energy narrowband terahertz
  generation,} {\protect\JournalTitle{Opt. Express}} \textbf{31}, 4041--4054
  (2023).

\bibitem{Eis}
H.~Eisele, A.~Rydberg, and G.~I. Haddad, \enquote{Recent advances in the
  performance of inp gunn devices and gaas tunnett diodes for the 100-300-ghz
  frequency range and above,} {\protect\JournalTitle{IEEE Transactions on
  Microwave Theory and Techniques}} \textbf{48}, 626--631 (2000).

\bibitem{mat}
A.~Crocker, H.~A. Gebbie, M.~F. Kimmitt, and L.~E.~S. Mathias,
  \enquote{Stimulated emission in the far infra-red,}
  {\protect\JournalTitle{Nature}} \textbf{201}, 250--251 (1964).

\bibitem{fatt}
C.~Fattinger and D.~Grischkowsky, \enquote{Terahertz beams,}
  {\protect\JournalTitle{Applied Physics Letters}} \textbf{54}, 490--492
  (1989).

\bibitem{zhang}
L.~Zhang, Y.~Jiang, W.~Lei, P.~Hu, J.~Guo, R.~Song, X.~Tang, G.~Ma, H.~Chen,
  and Y.~Wei, \enquote{A piecewise sine waveguide for terahertz traveling wave
  tube,} {\protect\JournalTitle{Scientific Reports}} \textbf{12} (2022).

\bibitem{xiaogen}
X.~Yi, H.~Zeng, S.~Gao, and C.~Qiu, \enquote{Design of an ultra-compact
  low-crosstalk sinusoidal silicon waveguide array for optical phased array,}
  {\protect\JournalTitle{Optics Express}} \textbf{28}, 37505--37513 (2020).

\bibitem{gratus2015}
J.~Gratus and M.~McCormack, \enquote{Spatially dispersive inhomogeneous
  electromagnetic media with periodic structure,}
  {\protect\JournalTitle{Journal of Optics}} \textbf{17} (2015).

\bibitem{boyd2018Cust}
T.~Boyd, J.~Gratus, P.~Kinsler, and R.~Letizia, \enquote{Customizing
  longitudinal electric field profiles using spatial dispersion in dielectric
  wire arrays,} {\protect\JournalTitle{Optics Express}} \textbf{26}, 2478--2494
  (2018).

\bibitem{boyd2018Mode}
T.~Boyd, J.~Gratus, P.~Kinsler, R.~Letizia, and R.~Seviour, \enquote{Mode
  profile shaping in wire media: Towards an experimental verification,}
  {\protect\JournalTitle{Applied Sciences (Switzerland)}} \textbf{8} (2018).

\bibitem{Bratman2000}
V.~L. Bratman, A.~W. Cross, G.~G. Denisov, W.~He, A.~D.~R. Phelps, K.~Ronald,
  S.~V. Samsonov, C.~G. Whyte, and A.~R. Young, \enquote{High-gain wide-band
  gyrotron traveling wave amplifier with a helically corrugated waveguide,}
  {\protect\JournalTitle{Physical Review Letters}} \textbf{84}, 2746--2749
  (2000).

\bibitem{Cook1998}
S.~J. Cooke and G.~G. Denisov, \enquote{Linear theory of a wide-band gyro-twt
  amplifier using spiral waveguide,} {\protect\JournalTitle{IEEE Transactions
  on Plasma Science}} \textbf{26}, 519--530 (1998).

\bibitem{Denisov2000}
G.~G. Denisov, V.~L. Bratman, A.~W. Cross, W.~He, A.~D.~R. Phelps, K.~Ronald,
  S.~V. Samsonov, and C.~G. Whyte, \enquote{Gyrotron traveling wave amplifier
  with a helical interaction waveguide,} {\protect\JournalTitle{Physical Review
  Letters}} \textbf{81}, 5680--5683 (1998).

\bibitem{nusinovich2004introduction}
G.~S. Nusinovich, \emph{Introduction to the Physics of Gyrotrons} (JHU Press,
  2004).

\bibitem{Nanni2015}
E.~A. Nanni, W.~R. Huang, K.~. Hong, K.~Ravi, A.~Fallahi, G.~Moriena, R.~J.
  Dwayne~Miller, and F.~X. Kärtner, \enquote{Terahertz-driven linear electron
  acceleration,} {\protect\JournalTitle{Nature Communications}} \textbf{6}
  (2015).

\bibitem{Zhao2019}
L.~Zhao, Z.~Wang, H.~Tang, R.~Wang, Y.~Cheng, C.~Lu, T.~Jiang, P.~Zhu, L.~Hu,
  W.~Song, H.~Wang, J.~Qiu, R.~Kostin, C.~Jing, S.~Antipov, P.~Wang, J.~Qi,
  Y.~Cheng, D.~Xiang, and J.~Zhang, \enquote{Terahertz oscilloscope for
  recording time information of ultrashort electron beams,}
  {\protect\JournalTitle{Physical Review Letters}} \textbf{122} (2019).

\bibitem{Wong2013}
L.~J. Wong, A.~Fallahi, and F.~X. Kärtner, \enquote{Compact electron
  acceleration and bunch compression in thz waveguides,}
  {\protect\JournalTitle{Optics Express}} \textbf{21}, 9792--9806 (2013).

\bibitem{Hibberd2020}
M.~T. Hibberd, A.~L. Healy, D.~S. Lake, V.~Georgiadis, E.~J.~H. Smith, O.~J.
  Finlay, T.~H. Pacey, J.~K. Jones, Y.~Saveliev, D.~A. Walsh, E.~W. Snedden,
  R.~B. Appleby, G.~Burt, D.~M. Graham, and S.~P. Jamison,
  \enquote{Acceleration of relativistic beams using laser-generated terahertz
  pulses,} {\protect\JournalTitle{Nature Photonics}} \textbf{14}, 755--759
  (2020). Cited By :47.

\bibitem{LuchiniMotzFEL}
P.~Luchini and H.~Motz, \emph{{Undulators and free electron lasers}} (Oxford
  University Press, 1990), pp. 106--112.

\bibitem{MadeyOrig}
J.~M.~J. Madey, \enquote{Relationship between mean radiated energy, mean
  squared radiated energy and spontaneous power spectrum in a power series
  expansion of the equations of motion in a free-electron laser,}
  {\protect\JournalTitle{Il Nuovo Cimento B Series 11}} \textbf{50}, 64--88
  (1979).

\bibitem{yan}
S.~Yan, \enquote{The gain calculation of media and electrostatic free-electron
  lasers by the madey theorem,} {\protect\JournalTitle{IEEE Journal of Quantum
  Electronics}} \textbf{23}, 1642--1645 (1987).

\bibitem{sev1}
Y.~S. Tan and R.~Seviour, \enquote{Wave energy amplification in a
  metamaterial-based traveling-wave structure,} {\protect\JournalTitle{EPL}}
  \textbf{87} (2009).

\bibitem{Eastham1973TheST}
M.~S.~P. Eastham, \emph{The spectral theory of periodic differential equations}
  (Scottish Academic Press, 1973).

\bibitem{ArscottMathiueEq}
F.M.Arscott, \emph{{Periodic differential equations}} (Pergamon, 1964), pp.
  124--127.

\bibitem{xfel}
J.~M.~J. Madey, \enquote{Stimulated emission of bremsstrahlung in a periodic
  magnetic field,} {\protect\JournalTitle{Journal of Applied Physics}}
  \textbf{42}, 1906--1913 (1971).

\end{thebibliography}


\begin{thebibliography}{}
\newcommand{\enquote}[1]{``#1''}

\end{thebibliography}

%%%%%%%%%%%%%%%%%%%%%%%%%%%%%%%%%%%%%%%%%%%%%%%%%%%%%%%%%%%%%%%%%%%%%%

\end{document}

\maketitle

%%%%%%%%%%%%%%%%%%%%%%%%%%%%%%%%%%%%%%%%%%%%%%%%%%%%%%%%%%%%%%%%%%%%%%

\section{Analysis of solutions to Mathieu's equation and origin of the zones}
\label{ch_AppMat}
%%%%%%%%%%%%%%%%%%%%%%%%%%%%%%
\begin{figure}
\begin{center}
% plot([Re(MathieuExponent(a, 0.1)), Im(MathieuExponent(a, 0.1)), Re(MathieuExponent(a, 0.5)), Im(MathieuExponent(a, 0.5)), Re(MathieuExponent(a, 0.9)), Im(MathieuExponent(a, 0.9))], a = 0 .. 6, color = [green, green, blue, blue, red, red], linestyle = [solid, dash], labels = ["", ""]);

\begin{tikzpicture}
\draw (0,0) node{
\includegraphics[width=0.5\textwidth]{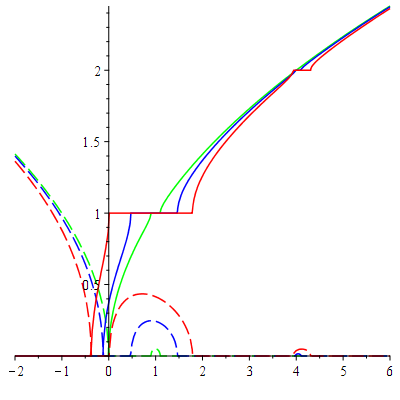}
};
\draw (1.2,-3.5) node {$a$} ;
\draw (-4.5,1.6) node {$\kscale(a,q)$} ;
\end{tikzpicture}
\caption{Plot of $\kscale(a,q)$, with $\Re(\kscale)$ solid and $\Im(\kscale)$
  dashed, and different $q$,
  $q=0.1$, (green), $q=0.5$ (blue) and $q=0.9$ (red). It is clear
  there are band gaps at $\kscale=1,2,\ldots$. }
\label{fig_MathieuExp}
\end{center}
\end{figure}
%%%%%%%%%%%%%%%%%%%%%%%%%%%%%%

In this section we show that the solutions to Mathieu's equation are invariant when $\kscale$ is replaced by $\kscale+2$ and $2-\kscale$. This is important for electron wave interaction as a single solution to the electromagnetic field will interact with electrons with many velocities, as seen in figure \ref{fig_Mat-Force_Disp}. For the consideration of the CIP in the main test we look as interaction in the 3rd zone when $2<\kscale<3$.

Mathieu's equation is an example of a Floquet equation,
with solution of the form given by (\ref{Mat_Floque}).
The Mathieu exponent $\kscale(a,q)$ can be found using matrix methods and is shown in figure \ref{fig_MathieuExp}.
This solution represents a mode with propagation constant $\kscale$
when $\kscale$ is real, while a complex $\kscale$ implies a propagation bandgap.

From equation (\ref{Mat_Floque}) we see that if $\kscale$ is a Mathieu exponent,
then so is $\kscale+2$ since 
%[
\begin{align*}
\psi_{a,q}(\zeta)
=
e^{i\kscale\zeta}\,\MatPF_{a,q,\kscale}(\zeta)
=
e^{i(\kscale+2)\zeta}\,e^{-2i\zeta}\MatPF_{a,q,\kscale}(\zeta)
=
e^{i(\kscale+2)\zeta}\,\MatPF_{a,q,\kscale+2}(\zeta)
\end{align*}
%]
hence
%[
\begin{align}
\MatPF_{a,q,\kscale+2}(\zeta) = e^{-2i\zeta}\MatPF_{a,q,\kscale}(\zeta)
\label{Mat_Floque_k+2}
\end{align}
%]

In addition, since Mathieu's equation (18) is a real
equation, then if $\psi_{a,q}(\zeta)$ is a solution, so is
$\cnj{\psi_{a,q}(\zeta)}$.  
%[
\begin{align*}
\cnj{\psi_{a,q}(\zeta)}
=
e^{-i\kscale\zeta}\,\cnj{\MatPF_{a,q,\kscale}(\zeta)}
=
e^{-i\kscale\zeta}\,\MatPF_{a,q,-\kscale}(\zeta)
\end{align*}
%]
hence
%[
\begin{align}
\MatPF_{a,q,-\kscale}(\zeta) = \cnj{\MatPF_{a,q,\kscale}(\zeta)}.
\label{Mat_Floque_-k}
\end{align}
%]

%%%%%%%%%%%%%%%%%%%%%%%%%%%%%%%%%%%%%%%%%%%%%%%%%%%%%%%%%%%%%%%%%%%%%%
\begin{figure}[htbp]
\centering\includegraphics[width=13cm]{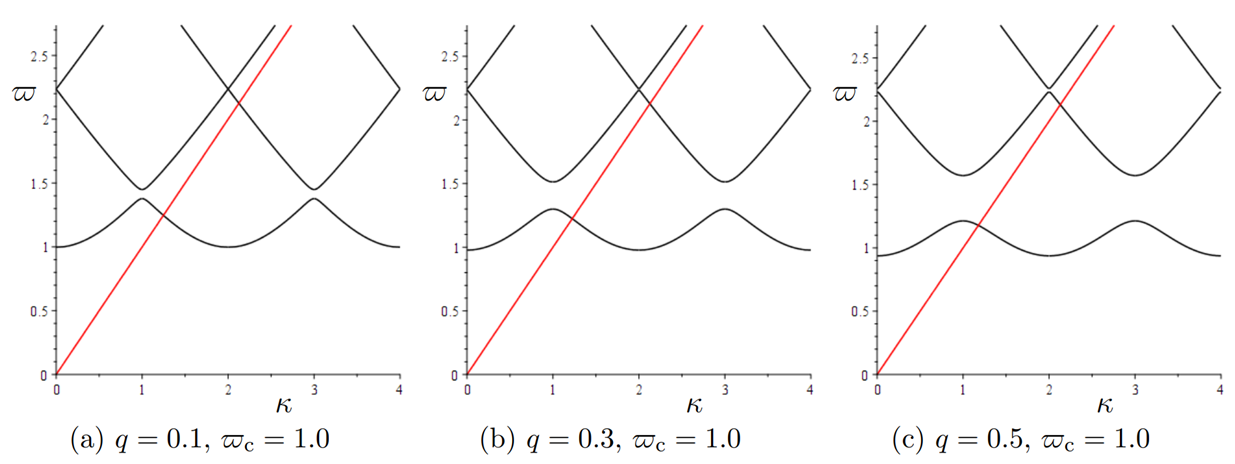}
\caption{Scaled dispersion relation (black). frequency $\fsc$ versus wavevector
  $\kscale$, for the different values of $q$ and fixed $\fscC$. Observe that
  as $q$ increases the band gap increases.
The red line is for $\fsc=\kscale$.
}
\label{fig_Disp_q}
\end{figure}

%%%%%%%%%%%%%%%%%%%%%%%%%%%%%%%%%%%%%%%%%%%%%%%%%%%%%%%%%%%%%%%%%%%%%%
\begin{figure}[htbp]
\centering\includegraphics[width=13cm]{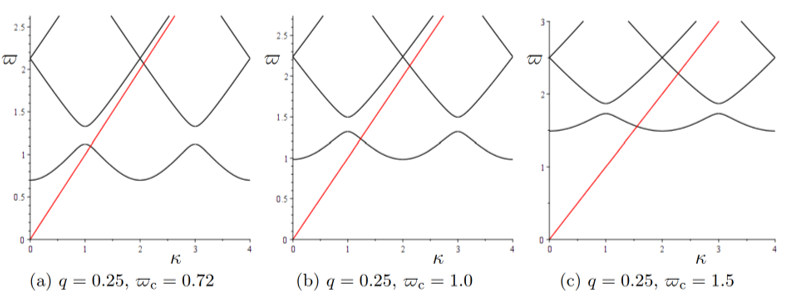}
\caption{Scaled dispersion relation (black). frequency $\fsc$ versus wavevector
  $\kscale$, for the different values of $\fscC$ and fixed $q$. Observe that
  as $\fscC$ increases the lowest modes increases.
The red line is for $\fsc=\kscale$.
}
\label{fig_Disp_omC}
\end{figure}

In figure \ref{fig_Disp_q} we show the effect of increasing $q$. As
can be seen this increases the band gap. In figure \ref{fig_Disp_omC}
we show the effect of increasing $\fscC$. In this case is pushes up
the lowest modes.
As these were analytic functions, generating these plots took less
than a few seconds, using MAPLE on a personal computer.
This enables us to find values which may be particularly useful, such as the CIP.

%%%%%%%%%%%%%%%%%%%%%%%%%%%%%%%%%%%%%%%%%%%%%%%%%%

%%%%%%%%%%%%%%%%%%%%%%%%%%%%%%%%%%%%%%%%%%%%%%%%%%%%%%%%%%%%%%%%%%%%%%
\section{Error analysis}
\label{ch_Appx_ErrBdd}

By substituting (\ref{Sol_B_flat}), (\ref{Sol_E}) into
Maxwell (\ref{Sol_Max_Apprx_Amp}) we calculate the error
$\ErrMax$ in Maxwell's equation
%[
\begin{equation}
\begin{aligned}
\ErrMax
&=
B_0\,\kappa_y\sin(\kappa_y\,y)
\big(
2\eta \kappa'_x(z)\,\sin(\,x)\,\phi'(\eta\,z)\,x
+
\kappa''_x(z)\,\sin(\kappa_x(z)\,x)\,\phi(\eta\,z)\,x
\\&\qquad\qquad
+
\kappa'_x(z)^2\,\cos(\kappa_x(z)\,x)\,\phi(\eta\,z)\,x^2
\big)\,\Vi
\\&\quad
+
B_0\, \cos(\kappa_y\,y)
\Big(
2\eta\,\kappa'_x(z)
\big(\sin(\kappa_x(z)\,x)+\kappa_x(z)\,\cos(\kappa_x(z)\,x)\,x\big)
\phi'(\eta\,z)
\\&\qquad\qquad
+
\big(
\kappa''_x(z)\sin(\kappa_x(z)\,x)
+
2\kappa'_x(z)^2\cos(\kappa_x(z)\,x)\,x
+
\kappa''_x(z)\kappa_x(z)\,\cos(\kappa_x(z)\,x)\,x
\\&\qquad\qquad
-
\kappa'_x(z)^2\kappa_x(z)\,\sin(\kappa_x(z)\,x)\,x^2
\big)
\phi(\eta\,z)
\Big)\,\Vj
\\&\quad
\end{aligned}
\label{Sol_ErrMax}
\end{equation}
%]

The boundary of the waveguide is 
%[
\begin{equation}
\begin{aligned}
\begin{cases}
y = \pm \Ly/2 &
\text{with normal unit vector $\Vj$ and tangent unit vectors
  $\Vi$ and $\Vk$.}
\\
x = \pm \Lx(z)/2 &
\text{\parbox{0.7\textwidth}
{with normal vector $(1+(\Lx')^2)^{-1/2}(\Vi-\Lx' \Vk)$ and tangent vectors
  $\Vj$ and $(1+(\Lx')^2)^{-1/2}(\Vk+\Lx' \Vi)$.}}
\end{cases}
\end{aligned}
\label{Sol_bdd}
\end{equation}
%]
We can now look at the boundary conditions
(\ref{Sol_Max_Bdd_BPerp}). 

With regards to $\VBft$, we see that along the $y = \pm \Ly/2$
%[
\begin{align*}
\VBft_\perp 
&=
\VBft_y \big|_{y=\Ly/2}
=
-
\big(B_0\, c^{-2}\,(-i \omega)\,
\kappa_x(z)\,\sin(\kappa_x(z)\,x)\,\cos(\kappa_y\,y)\big)\big|_{y=\Ly/2}
=0
\end{align*}
%]
while along $x = \pm \Lx(z)/2$ we have
%[
\begin{align*}
\VBft_\perp 
&=
\big((1+(\Lx')^2)^{-1/2} \,\VBft\cdot (\Vi-\Lx' \Vk)\big) 
\big|_{x = \pm \Lx(z)/2}
\\&=
\big(
B_0\, c^{-2}\,(-i \omega)\,\phi(\eta\,z)\,
\kappa_y\cos(\kappa_x(z)\,x)\,\sin(\kappa_y\,y)
\big)\big|_{x = \pm \Lx(z)/2}
=0
\end{align*}
%]
Hence (\ref{Sol_Max_Bdd_BPerp}) holds.

With regards to $\VEft$ we can calculate $\ErrBdd$ from
(\ref{Sol_Max_Apprx_Amp}). This is given for $y = \pm \Ly/2$ by
%[
\begin{align*}
\VEft\cdot\Vi
\big|_{y = \pm \Ly/2}
%\\
&
=
B_0
\Big(\big(\kappa'_x(z)\sin(\kappa_x(z)\,x)+\kappa'_x(z)\kappa_x(z)\,\cos(\kappa_x(z)\,x)\big)
\phi(\eta\,z)
\\&\qquad\qquad\qquad\qquad\qquad\qquad 
+
\eta\kappa_x(z)\,\sin(\kappa_x(z)\,x)\,\phi'(\eta\,z)
\Big)
\cos(\kappa_y\,y)
\Big|_{y = \pm \Ly/2}
%\\&
=0
\end{align*}
%]
and
%[
\begin{align*}
\VEft\cdot\Vk
\big|_{y = \pm \Ly/2}
&=
-  B_0
\Big(
(\kappa_x(z)^2+\kappa_y^2)
\,\cos(\kappa_x(z)\,x)\,\cos(\kappa_y\,y)\phi(\eta\,z) 
\Big)|_{y = \pm \Ly/2}
=0
\end{align*}
%]
Hence along the flat edge $\ErrBdd=0$.

Along the corrugations $x = \pm \Lx(z)/2$ we have
$\ErrBdd=\Vzero$ along flat side and
%[
\begin{equation}
\begin{aligned}
\ErrBdd
&=
- (-1)^{(p-1)/2} B_0\, \px\,
\big(
(2\kappa_x(z))^{-1}\,\pi\,\sin(\kappa_y\,y)
\kappa_y\, 
\kappa'_x(z)\,\phi(\eta\,z)\,
\Vj
\\&
\pm (-1)^{(p-1)/2} (1+(\Lx')^2)^{-1}\,
\Lx'\,\cos(\kappa_y\,y)
\big(
\kappa'_x(z)\,
\phi(\eta\,z)
+
\eta\kappa_x(z)\,\phi'(\eta\,z)
\big) 
\cos(\kappa_y\,y)\,(\Vk+\Lx' \Vi)\big)
\label{Sol_Bdd_ErrBdd}
\end{aligned}
\end{equation}
%]
along corrugations.

%%%%%%%%%%%%%%%%%%%%%%%%%%%%%%%%%%%%%%%%%%%%%%%%%%%%%%%%%%%%%%%%%%%%%%

Looking at the errors $\ErrMax$ and $\ErrBdd$ we see from
(\ref{Sol_ErrMax}) and (\ref{Sol_Bdd_ErrBdd}) that there is an overall
factor of either $\kappa'_x(z)$, $\kappa''_x(z)$ or 
$\Lx'(z)$. From (\ref{Sol_lambda_z}) we
have 
%[
\begin{align}
\kappa'_x(z)(z) = - \pi\,\Lx^{-2}\,\Lx'(z)
\qquadand
\kappa''_x(z)(z) = 2 \pi\,\Lx^{-3}\,\big(\Lx'(z)\big)^2 - \pi\,\Lx^{-2}\,\Lx''(z)
\label{Sol_Err_alpha'}
\end{align}
%]
Hence all terms have a factor of either $\Lx'$ or $\Lx''$.  Clearly
(\ref{Sol_B_flat}), (\ref{Sol_E}) is a good approximation in a
corrugation waveguide where $\Lx'$ and $\Lx''$ are small. For
example, where there is a smooth and gentle corrugation. 

In order to improve the accuracy of the solution, one could consider
an expansion in small parameter $\delta_1$ where 
$\delta_1\approx\max(\Lx',\Lx\,\Lx'')$
%[
\begin{align}
\VEft_{\textup{exact}} = \VEft_{(0)} + \delta_1 \VEft_{(1)} + \cdots
\qquadand
\VBft_{\textup{exact}} = \VBft_{(0)} + \delta_1 \VBft_{(1)} + \cdots
\label{Sol_Err_expansion}
\end{align}
%]
where $\VEft_{(0)}$ and $\VBft_{(0)}$ are given by (\ref{Sol_E}) and
(\ref{Sol_B_flat}). Substituting (\ref{Sol_Err_expansion}) into
Maxwell (\ref{Sol_Max_Apprx_Amp})-(\ref{Sol_Max_Bdd_BPerp}) will give the
differential equations for corrections $\VEft_{(1)}$, $\VBft_{(1)}$, etc.

%%%%%%%%%%%%%%%%%%%%%%%%%%%%%%%%%%%%%%%%%%%%%%%%%%%%%%%%%%%%%%%%%%%%%%

\begin{figure}[htbp]
\centering
\setlength{\jglength}{7cm}
\begin{tikzpicture}[x=\jglength,y=\jglength]
\draw (.47,.47) node{\includegraphics[height=1.08\jglength]{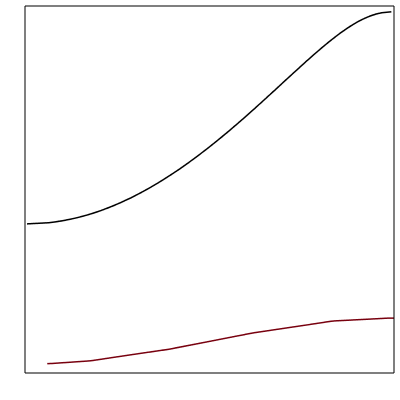}} ;
\foreach \x in {0,0.1,0.2,0.3,0.4,0.5,0.6,0.7,0.8,0.9,1}
 {\draw[thick] (\x,0) -- +(0,-.01)  node[below]{\x};} ;
\draw(0.5,-.05) node[below]{$\kscale$} ;
\foreach \x in {1.0,1.1,1.2,1.3,1.4}
 {\draw[thick] (0,{(\x-0.99)/(-0.99+1.43)}) -- +(-.02,0)  node[left]{\x};} ;
\draw(-.05,0.55) node[left]{$\fsc$} ;
% \foreach \x in {240,260,...,320}
%  {\draw[thick] (1,{(\x-235)/(-235+325)}) -- +(.02,0)  node[right]{\x};} ;
% \draw(1.05,0.6) node[right]{$\omega(\text{GHz})$} ;
% \foreach \x in {0,200,400,600}
%  {\draw[thick] ({(\x)/(790)},1) -- +(0,.02)  node[above]{\x};} ;
% \draw(0.55,1.1) node[right]{$k(\meters^{-1})$} ;
\end{tikzpicture}
\qquad
\raisebox{1em}{\includegraphics[width=0.2\textwidth]{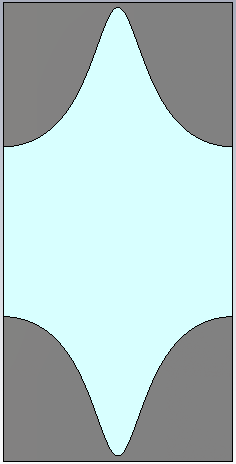}}
%\qquad
%\includegraphics[width=0.6\textwidth]{q_0_39_xc_1_2_Structure1.png}

\caption{Numerical (red) versus predicted (black) dispersion
  plot. Here $q=0.39$, $\fscC=1.2$, $\Lz=4\mmeters$,
  $\Lnul=8.0\mmeters$ and $\Ly=6.3\mmeters$. This is outside the domain
  of validity of the approximation and we see that the two and no
  longer in good agreement.}
\label{fig_Siml_DR2_bad}
\end{figure}

 From (\ref{Match_Lx_L0}) we have
%[
\begin{align*}
\kappa_x(z)^2 =  
\pi^2 \,\Lnul^{-2} + 
2\pi^2 \,\Lz^{-2}\, q\,\cos(2\,\pi\,\Lz^{-1}\,z) 
\end{align*}
%]
Hence
%[
\begin{gather}
\kappa'_x(z)
=
-\frac{2q\pi^3}{\Lz^3\,\kappa_x(z)}\sin(2\,\pi\,\Lz^{-1}\,z) 
\quadand
\notag
\\
\kappa''_x(z)
=
-
\frac{4q\pi^4}{\Lz^4\,\kappa_x(z)}\cos(2\,\pi\,\Lz^{-1}\,z)
-
\frac{4q^2\pi^4}{\Lz^6\,\kappa_x(z)^3}\sin^2(2\,\pi\,\Lz^{-1}\,z)
\label{Match_Err}
\end{gather}
%]
Hence for a fixed $q$ and $\Lnul$ we see that the errors $\ErrMax$ and
$\ErrBdd$ goes to zero as $\Lz^{-3}\to\infty$.

It is important to know how the errors go to zero as a function of
dimensionless parameters. Let
%[
\begin{align}
\delta = \frac{L_0}{L_z}
\label{Accu_def_delta}
\end{align}
%]
Then we can calculate all the expressions in terms of $\delta$ and
$q$. To simplify the analysis we assume $\Lx\approx\Lnul$ and
$x\approx\Lnul$. 

We use the notation $A\sim\delta^r$ to mean that the quantity $A$ go
to zero at a rate approximately equal to $\delta^r$. Considering all
the interesting terms we have
%[
\begin{align}
\Lx'(z) \sim \delta^3 q
,\quad
\Lnul\kappa_x(z) \sim \delta^0
,\quad
\Lnul^2 \kappa'_x(z) \sim \delta^3 q
,\quad
\Lnul^3 \kappa''_x(z) \sim \max(\delta^4 q,\delta^6 q^2)
,\quad
\fscC \sim \delta^{-1}
\label{Accu_Err_param}
\end{align}
%]
while the errors $\ErrMax$ and $\ErrBdd$ go to zero as
%[
\begin{align}
\Lnul^3\,\ErrMax \sim \max(\delta^4 q,\delta^6 q^2)
\qquadand
\Lnul^3\ErrBdd \sim \max(\delta^3 q,\delta^4 q,\delta^6\,q^2)
\label{Accu_Errors}
\end{align}
%]
From these we can establish which parameters will result in a good
approximation. We can see we can either choose $\delta$ to be small,
in which case $\fscC$ will be large. Alternatively we can set $q$ to
be small. In the examples we have we see that a good approximation is
achieved when $\delta\approx 1$ and $q\approx 0.1$.

%%%%%%%%%%%%%%%%%%%%%%%%%%%%%%%%%%%%%%%%%%%%%%%%%%%%%%%%%%%%%%%%%%%%%%

\section{The impossibility of subluminal modes in the first  zone}
\label{ch_Imp_First}

\begin{figure}[htbp]
\begin{center}
\includegraphics[width=0.62\textwidth]{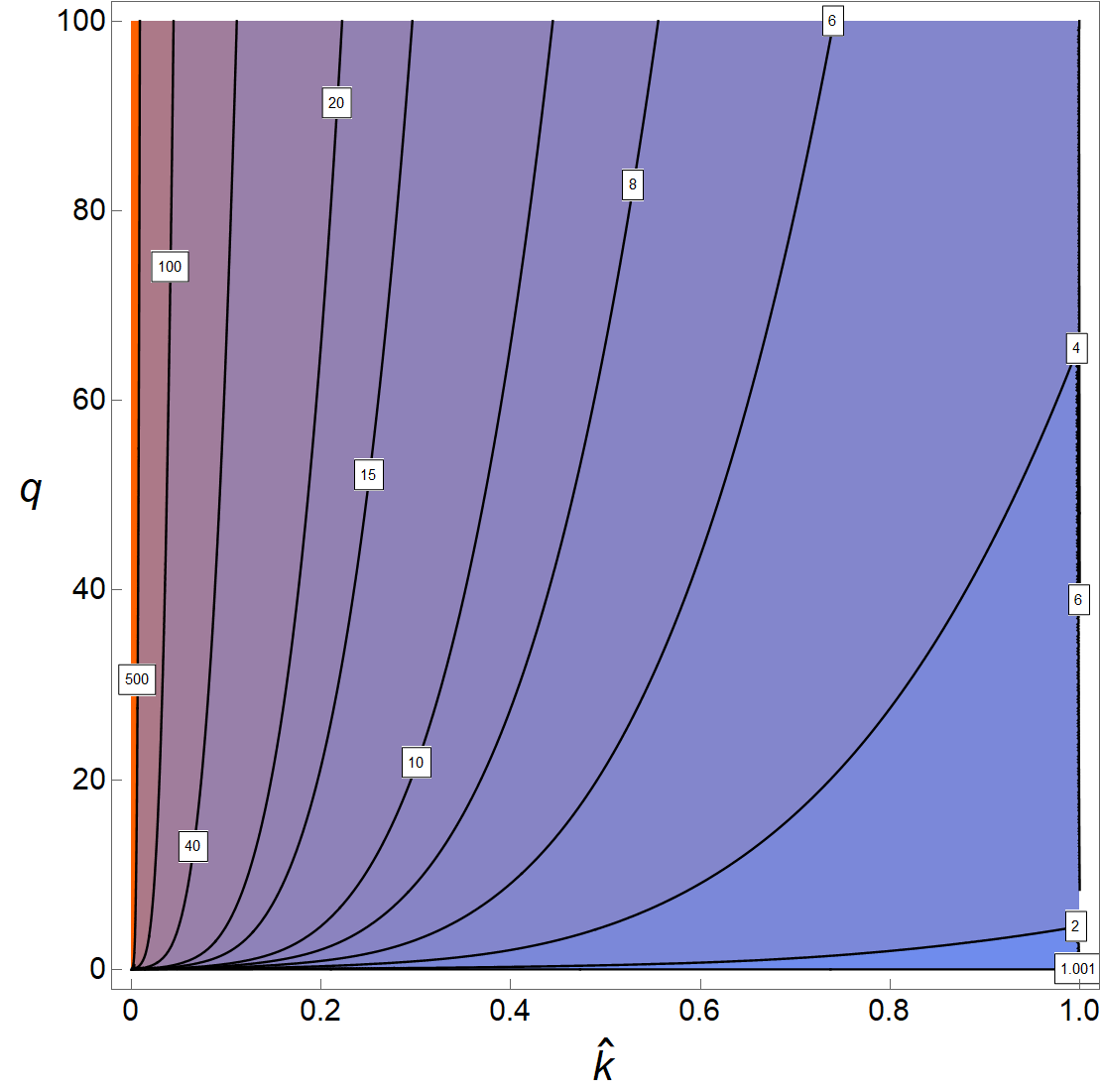}
\caption{\centering Graphically showing values of the left hand side of (\ref{fsc_neq3}) inequality. This shows that the values are all greater than 1.}
\label{fig_NEQ}
\end{center}
\end{figure}

In figures \ref{fig_Mat-Force_Disp}, \ref{fig_Disp_q} and \ref{fig_Disp_omC} we see that the intersection of the lightline with all the dispersion curve is always for $\kscale>1$, i.e. outside the first  zone. We see here why this must be the case.
For physical
solutions $\Lx(z)>0$, we require
%[
\begin{align}
2\,q < \Lnul^{-2}\,\Lz^{2}
\label{Match_constraint}
\end{align}
%]
hence we have the constraint
%[
\begin{align}
\fscC^2 > 2 q
\label{Match_constraint_scaled}
\end{align}
%]
Scaled angular frequency $\fsc$ is 
\begin{align}
\fsc^2 = a + \fscC^2
\label{fsc_remdef}
\end{align}
Thus using (\ref{Match_constraint_scaled}) yields
\begin{align}
a + 2\fscC^2 > a + 2q  \quad \Rightarrow \quad \fsc^2 > a + 2q \quad \Rightarrow \quad \fsc > \sqrt{a + 2q}
\label{fsc_neq1}
\end{align}
Reformulating the last inequality so that phase velocity $\frac{\fsc}{\kscale}$ is in the left hand side of the inequality (for $\kscale\neq0$)
\begin{align}
\frac{\fsc}{\kscale}>\frac{\sqrt{a+2q}}{\kscale}
\label{fsc_neq2}
\end{align}
Mathiue parameter $a$ is known function of $\kscale$ and $q$. Numerically plotting (Fig. \ref{fig_NEQ}) values of a function $[a\left(\kscale,q\right)+2q]^{1/2}/\kscale$ shows that for the wavenumbers in the first zone, i.e. $0<\kscale\leq1$ the function values are higher than 1.
\begin{equation}
\frac{\sqrt{a\left(\kscale,q\right)+2q}}{\kscale}>1
\label{fsc_neq3}
\end{equation}
Thus $\fsc/\kscale>1$ and hence the region $\kscale<1$ is above the lightline.

%%%%%%%%%%%%%%%%%%%%%%%%%%%%%%
%%%%%%%%%%%%%%%%%%%%%%%%%%%%%%
% In figure \ref{fig_Disp_omC} there is an increase in $\fscC$ for the same $q$. We see there is a maximum value of $\fscC$ so that subluminal modes exist in the first  zone ($\kscale<1$).
% This is given by $\fscCsubml(q)$. Thus for
% subluminal first  zone modes we need $\fscC<\fscCsubml(q)$. 
% Looking at the solutions to Mathieu's equation, numerically, we can see that 
% we see that for $0<q<1$, $\fscCsubml^2(q) \approx q$
% and that definitely $\fscCsubml^2(q)<2q$ for $0<q<100$.
% This contradicts the constraint (\ref{Match_constraint_scaled}), and
% hence there is a
% contradiction and there are no subluminal modes in the first 
% zone.

% \color{red}

% Is it just testing following neq for any q<1 and $0<\kscale\leq1$?

% \begin{equation}
%    \frac{\fsc\left(\kscale,q\right)^2}{\kscale^2}>\frac{a\left(\kscale,q\right)+2q}{\kscale^2}>1
% \end{equation}

% \color{black}

%%%%%%%%%%%%%%%%%%%%%%%%%%%%%%%%%%%%%%%%%%%%%%%%%%%%%%%%%%%%%%%%%%%%%%
\section{Other waveguide modes}
\label{ch_OtherModes}

The approach detailed in section \ref{ch_Sol} of the main text  can be used to find equivalent approximate solution for the other waveguide modes. For the case of the $\TM_{\px\py}$ where $\px$ or $\py$ is even, we can start with eq. (\ref{Sol_B_flat}) but replace $\cos(\kappa_x(z)\,x)\leftrightarrow\sin(\kappa_x(z)\,x)$ and $\cos(\kappa_y\,y)\leftrightarrow\sin(\kappa_y\,y)$ where appropriate.

For the TE modes we start by replacing (\ref{Sol_B_flat}) with the
equivalent equation for the transverse electric field. In this case (\ref{Sol_Max_Apprx_NoMon})-(\ref{Sol_Max_Apprx_Amp}) would be replaced by
%[
\begin{align}
\nabla\cdot\VEft=0,\quad
\nabla\cdot\VBft=0,\quad
\nabla\times\VEft - i\omega{\VBft} = 0
\qquadand
\nabla\times\VBft + i\omega\,c^{-2}{\VEft} = \ErrMax
\label{SD_ME}
\end{align}
%]
and there may be error terms in the boundary for the electric and magnetic
fields.
%%%%%%%%%%%%%%%%%%%%%%%%%%%%%%%%%%%%%%%%%%%%%%%%%%%%%%%%%%%%%%